\documentclass{jfm}
\usepackage{graphicx}
\usepackage{epstopdf, epsfig}
\usepackage{amsmath}
\usepackage{mathrsfs} 
\usepackage{minitoc}
\usepackage[position=top]{subfig}
\usepackage{tikz}
\usepackage{pgfplots}
\usepackage{url}
\usetikzlibrary{arrows.meta}

\renewcommand{\vec}[1]{\mbox{\boldmath $ #1 $}}

\renewcommand{\vec}[1]{\mbox{\boldmath $ #1 $}}

\newcommand{\dprime}{{\prime\prime}}
\newcommand{\ii}{\mathrm i}
\newcommand{\ee}{\mathrm e}

\shorttitle{Transition in rough solitary wave boundary layers}
\shortauthor{A. \"Onder, P. L.-F. Liu}

\title{Receptivity and transition in a wave boundary layer over random bottom topography  }

\author  {Asim \"Onder \aff{1}
  \corresp{\email{asim.onder@gmail.com}},
Philip L.-F. Liu \aff{1,2,3}
  }

\affiliation{ 
\aff{1 }
Department of Civil and Environmental Engineering, National University of Singapore, Singapore 117576, Singapore
\aff{2}
School of Civil and Environmental Engineering, Cornell University, Ithaca, NY 14850, USA
\aff{3}
Institute of Hydrological and Oceanic Sciences, National Central University, Jhongli, Taoyuan, 320, Taiwan}

\begin{document}

\maketitle

\begin{abstract}
Direct numerical simulations are conducted to study the receptivity and transition mechanisms in a solitary wave boundary layer developing over randomly organized wave-like bottom topography. The boundary layer flow shows a selective response to broadband perturbations from the bottom, and develops streamwise-elongated streaks.  When the streaks reach high amplitudes,  they indirectly amplify streamwise-elongated vortices through modulating small-scale fluctuations and pressure fields. These stronger vortices in turn stir the boundary layer more effectively and further intensify streaks via the lift-up mechanism.  This nonlinear feedback loop increases the sensitivity of the boundary layer to the roughness level and yields dramatic variations among cases sharing the same Reynolds number with differing roughness height. Three different local breakdown scenarios are observed depending on the amplitude of the streaks: (i)  two-dimensional wave instabilities in the regions with weak streaks; (ii) inner shear-layer instabilities in the regions with moderate-amplitude streaks; (iii)  rapidly growing outer shear-layer instabilities in the regions with highly elevated high-amplitude streaks. Inner instabilities have the slowest growth rate among all transition paths, which confirms the previous predictions on the stabilising role of moderate-amplitude streaks (\"Onder \& Liu, \textit{J. Fluid. Mech.}, vol. 896, 2020, A20).

\end{abstract}


\section{Introduction}
Surface gravity waves in shallow waters often travel over random bottom topography composed of disorganised bedforms or coarse sediments, e.g., gravel or sand.  These small-scale features act as hydrodynamic roughness in the wave boundary layers developing over them.  While in fully-developed turbulent wave boundary layers the effect of roughness is usually well parameterised using classical concepts, e.g. Nikuradse roughness and logarithmic velocity profile, there are relatively few studies addressing the transitional regimes beneath mild waves.  Such random topography-induced transition is complicated and its building steps are not well understood. 
Using direct numerical simulations (DNS), the present work investigates the mechanics of boundary layer transition over a random bed topography beneath a solitary wave, which can be viewed as a simple reproducible prototype for long regular waves in the shoaling zone (e.g. \cite{munk1949solitary}). The primary objective of this study is to establish direct links between topography and precursor structures of transition and subsequent transition modes.  The bed is modelled as randomly organized wave-like corrugations and its geometry is well resolved by conducting DNS on a transformed coordinate system.  

 \begin{figure}
\begin{center}
\includegraphics{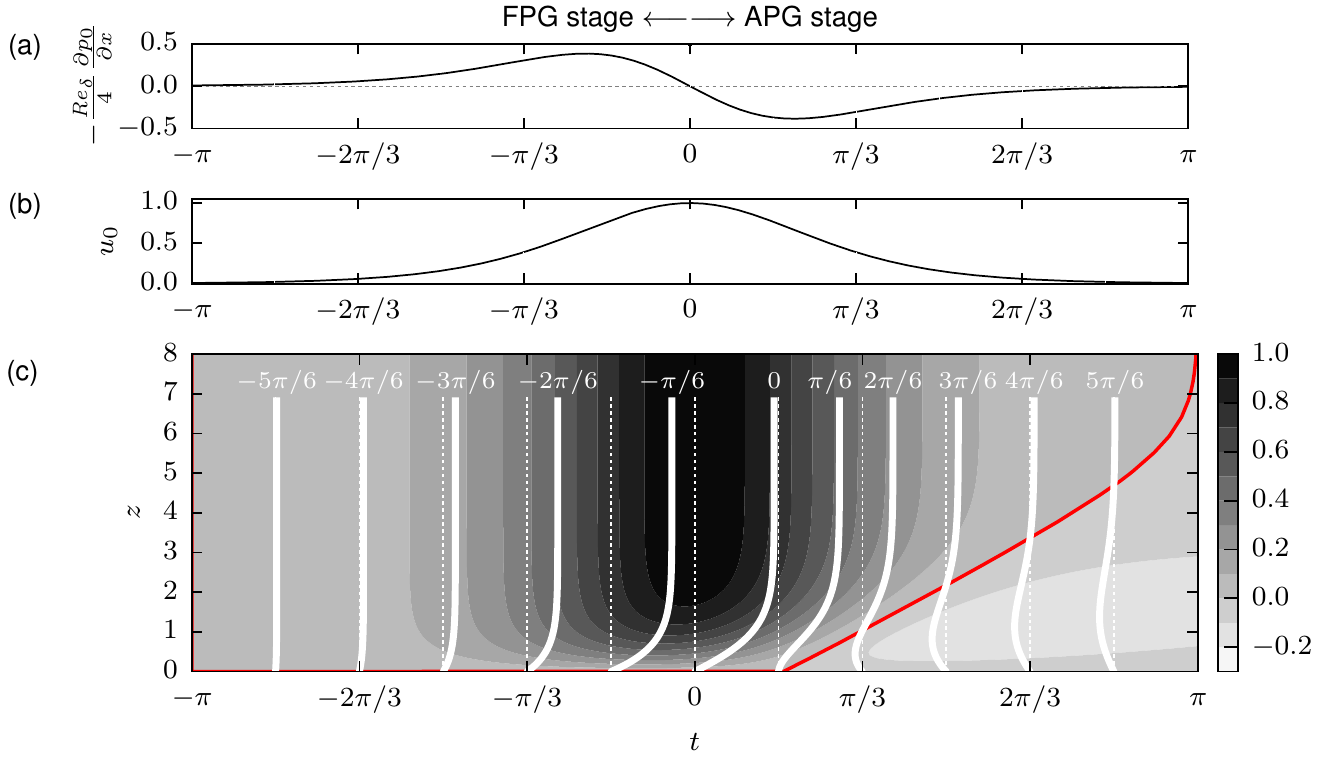}
\caption{ \label{fig:uMean} Temporal variation of the flow fields at a fixed location beneath a passing solitary wave. (\textit a) Pressure gradient. Two stages are defined: (i) favourable pressure gradient (FPG) stage ($t<0$); (ii) adverse pressure gradients (APG) stage ($t>0$). ($b$) Irrotational (free-stream) velocity above the solitary wave boundary layer (SWBL). (c) Velocity $u$ in the laminar SWBL. Vertical profiles of $u$ at phases $t=\{-5\pi/6,-4/\pi6,\ldots,5\pi/6\}$ are overlaid on contours. Red contourline is the level $u=0$. The definitions for free-stream fields and normalizations for lengths, velocities, pressure and time are introduced in \S~\ref{sec:conf}.}
\end{center}
\end{figure}

Solitary wave is a symmetric long wave with a single crest. It imposes an approximately constant horizontal velocity across the water column. A given point beneath an approaching wave experiences successive stages of accelerating and decelerating onshore velocities (figure~\ref{fig:uMean}$b$) driven by favourable and adverse pressure gradients (FPG and APG), cf. figure~\ref{fig:uMean}($a$). Unlike the irrotational flow above, the near-bed velocity in the boundary layer begins to decelerate at the end of the FPG stage and eventually  reverses its direction at the beginning of the APG stage \citep{LIU:2004jo,LIU:2007dv}, cf.~figure~\ref{fig:uMean}($c$). In this process, the adverse pressure gradient and frictional forces give rise to inflectional velocity profiles rendering the flow linearly unstable above critical wave amplitudes \citep{Blondeaux:2012ei,Sadek:2015jm}. As a result, two-dimensional instability waves can develop and grow into coherent spanwise vortex rollers with regular spacing \citep{Sumer:2010ce}. For higher wave amplitudes, these coherent vortices themselves are unstable and break into small-scale turbulence \citep{Vittori:2008gv,OZDEMIR:2013bu}. \cite{scandura13} showed in a two-dimensional numerical setting that the instability waves and coherent vortices can also be generated by wall imperfections of small amplitude.

\begin{figure}
\begin{center}
\pgfmathsetmacro{\dx}{4.4}
\pgfmathsetmacro{\dy}{-2.90}
\pgfmathsetmacro{\yT}{1.4}
\pgfmathsetmacro{\yA}{0.8}
\begin{tikzpicture}[declare function={swave(\x)=(1/(cosh(\x)^2);}]
\node (f1) at (\dx*0,0) {\includegraphics[width=.31\textwidth]{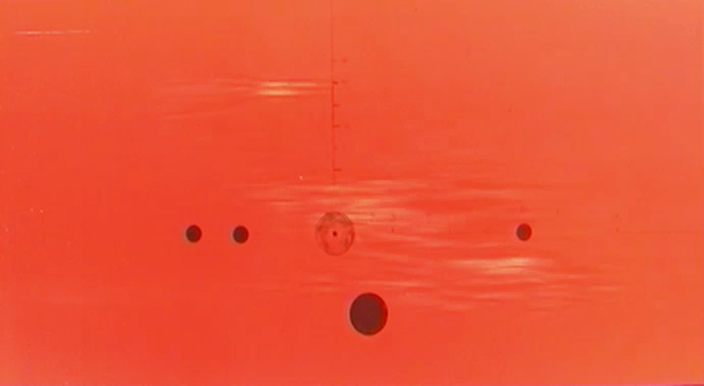} }+ (0,\yT) node[font=\small] {(a) $t=0.2$};
\draw [line width=0.5pt,-Triangle] (\dx*0-2,\yA) -- +({swave(0.2)},0);
\draw [line width=0.5pt,-Triangle,color=red] (\dx*0-2,\yA+0.6) -- +({swave(0.)},0);
\node (f2) at (\dx,0)  {\includegraphics[width=.31\textwidth]{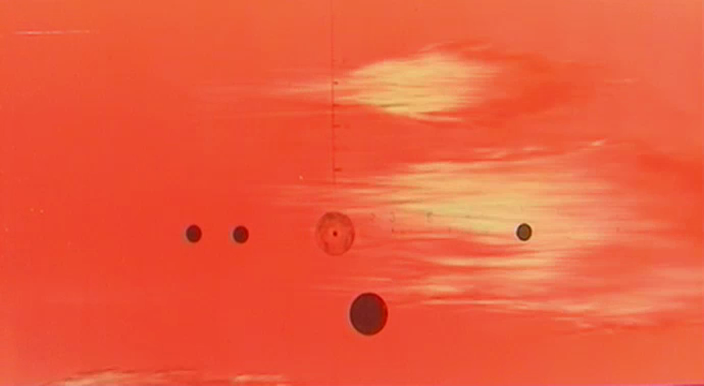}}+ (\dx,\yT) node[font=\small] {(b) $t=0.46$};
\draw [line width=0.5pt,-Triangle] (\dx-2,\yA) -- +({swave(0.46)},0);
\node (f3) at (2*\dx,0)   {\includegraphics[width=.31\textwidth]{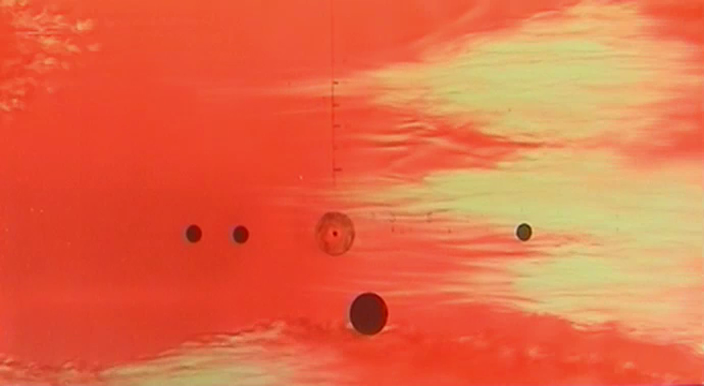}}+ (2*\dx,\yT) node[font=\small] {(c) $t=0.72$};
\draw [line width=0.5pt,-Triangle] (2*\dx-2,\yA) -- +({swave(0.72)},0);

\node (f4) at (\dx*0,\dy) {\includegraphics[width=.31\textwidth]{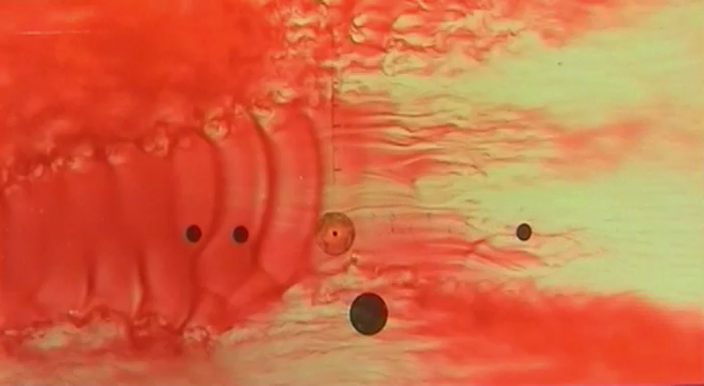} }+ (0,\dy+\yT) node[font=\small] {(d) $t=1.07$};
\draw [line width=0.5pt,-Triangle] (0*\dx*0-2,\dy+\yA) -- +({swave(1.07)},0);
\node (f5) at (\dx,\dy)  {\includegraphics[width=.31\textwidth]{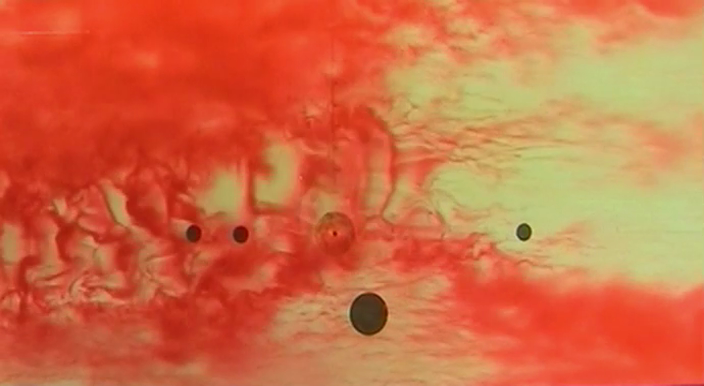}}+ (\dx,\dy+\yT) node[font=\small] {(e) $t=1.24$};
\draw [line width=0.5pt,-Triangle] (1*\dx-2,\dy+\yA) -- +({swave(1.24)},0);
\node (f6) at (2*\dx,\dy)   {\includegraphics[width=.31\textwidth]{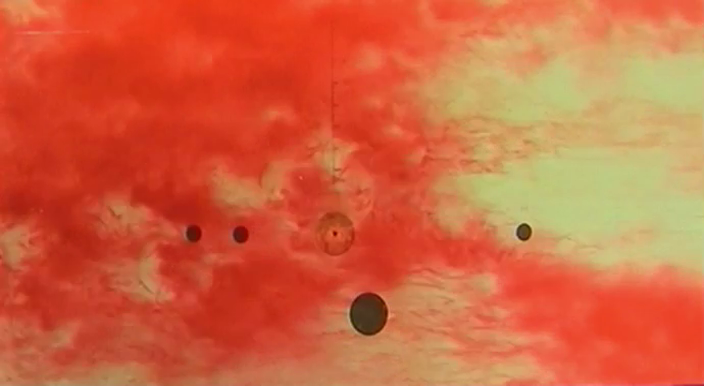}}+ (2*\dx,\dy+\yT) node[font=\small] {(f) $t=1.47$};
\draw [line width=0.5pt,-Triangle] (2*\dx-2,\dy+\yA) -- +({swave(1.47)},0);
\end{tikzpicture}

\caption{\label{fig:sumer} A sequence of video frames illustrating the transition modes in a solitary wave boundary layer at $\Rey_\delta=1483$: ($a$): streamwise streaks; ($b$--$f$): turbulent spots; ($d$--$f$): spanwise vortex rollers. The frames are extracted from the supplementary movie~3 in \cite{Sumer:2010ce} and reproduced with a permission granted from Cambridge University Press. Black arrows demonstrate the decelerating free-stream velocity. Red arrow in ($a$) indicates maximum free-stream velocity. Time is normalized using the wave frequency, i.e., $t=t^*\omega^*$ (cf. \S~\ref{sec:conf}).  }
\end{center}
\end{figure}
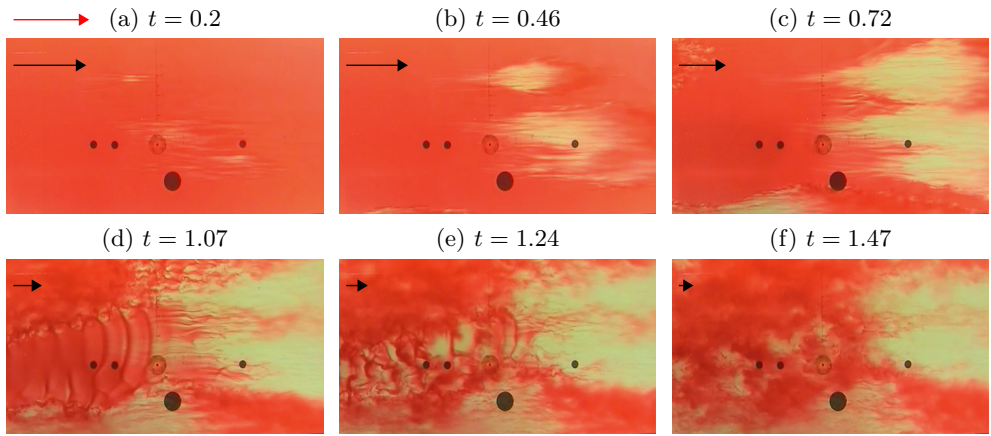

The orderly two-dimensional path to transition is often accompanied or ``bypassed" by transitional features of more disorganised stochastic nature, i.e., turbulent spots. \cite{Sumer:2010ce}  studied solitary wave boundary layer (SWBL) in an oscillatory water tunnel and turbulent spots were the first turbulent features emerging in such a flow. They were initially observed after the flow reversal in the APG stage. With increasing Reynolds number (to be defined in \S~\ref{sec:conf}) the spots were nucleated in earlier phases. A mixed transition scenario is demonstrated in figure~\ref{fig:sumer} using a sequence of video frames from APG stage (cf. supplementary movie~3 in \cite{Sumer:2010ce}). Turbulent spots emerge at early APG stage and starts to grow, cf. figures~\ref{fig:sumer}($a$--$c$). Before they spread everywhere in the boundary layer, coherent vortex rollers spontaneously emerge in the laminar regions surrounding the spots (figure~\ref{fig:sumer}$d$). The rollers quickly become unstable (figure~\ref{fig:sumer}$e$) and broke into smaller scales, which completes the transition to turbulence (figure~\ref{fig:sumer}$f$). 

Compared to the linear stability theory for instability waves, the theory behind the onset of turbulent spots is much less established and is largely phenomenological \citep{DURBIN:2017dn}. This stochastic transition path is initiated by the receptivity of the flow to finite-amplitude external perturbations such as bottom roughness or free-stream turbulence.  The boundary layer flow amplifies these perturbations and develops streamwise elongated regions of streamwise velocity fluctuations, termed as streaks. The early stages of streak amplification is usually explained mathematically by linear non-modal growth theory  \citep{butlerPOF92,trefethen1993hydrodynamic}. Physically, the streaks are generated by streamwise-oriented vortices stirring the streamwise momentum across the boundary layer, an effect knows as lift-up mechanism \citep{landahlJFM80}. Once the streaks reach high amplitudes, the boundary layer becomes strongly corrugated along its span and each streak hosts inflectional velocity profiles across vertical and spanwise cross-sections. As a result,  streaks become susceptible to inviscid instabilities \citep{Andersson:2001dm,cossu2002stabilization}. The most energetic streaks locally broke down due to these secondary instabilities, and formation of turbulent spots begins \citep{Vaughan:2011ho}. Several streaks prior to formation of turbulent spots can be seen in figure~\ref{fig:sumer}a. 

Unlike oscillatory boundary layers (e.g. \cite{carstensen2012note, mazzuoli2016transition,biau2016transient}), only a few studies focused to date on the stability characteristics of SWBLs in the presence of finite-amplitude perturbations. \cite{Verschaeve:2017fa} studied linear nonnormal growth in the SWBL. They showed that initial perturbations in the form of streamwise-constant counter-rotating vortex pairs can strongly amplify streamwise-constant streaks in the FPG stage with a maximum growth proportional to the square of Reynolds number. Later in the APG stage, the nonnormal growth of streaks are dominated by the nonnormal growth of two-dimensional instability waves, which grow exponentially in Reynolds number. \cite{onder_liu_2020} modelled external perturbations as distributed body forces and analyzed the receptivity of SWBL to these perturbations in a linear framework. They also identified streamwise-constant vortices as the most effective perturbations to generate streaks. They further deployed these optimal perturbations into nonlinear governing equations and analyzed the stability of the SWBL for various perturbation magnitudes. The resulting streaks have been found to play a dual role in the boundary layer stability. Low-to-moderate amplitude streaks have a dampening effect and delay the transition in the APG stage. In contrast, if the streaks are strongly amplified and elevated deep into the free stream, they can develop sinuous oscillations and initiate bypass-transition scenario already in the FPG stage. 

In the present work, we focus on a natural bypass-transition scenario, in which a solitary wave passes over random bottom topography containing wave-like undulations of finite amplitude. Particular attention is paid to the receptivity stage, where broadband perturbations introduced by irregular bottom roughness is filtered by SWBL and converted into energetic streamwise streaks. The linear and nonlinear stagess of the phenomenon are identified with special emphasis on dynamic feedback mechanisms between streamwise streaks and vortices. In the last step, the various transition paths to turbulence are illustrated for different roughness heights. 

The paper is organized as follows. First, the SWBL model along with the bottom topography function will be introduced in \S\ref{sec:conf}. Subsequently, numerical details of direct numerical simulations will be presented in  \S\ref{sec:numerics}. The analysis of results consists of two parts. In \S\ref{sec:streaks}, we will first focus on the boundary-layer response to bottom perturbations. The generation of streaks are analyzed in detail in this section.  Subsequently, in \S\ref{sec:breakdown}, various transition scenarios will be investigated. Finally, conclusions are summed up in \S\ref{sec:conclusion}.

\section{Methodology}
\subsection{Flow configuration}\label{sec:conf}
 We consider a SWBL model, in which streamwise scales are much larger than vertical scales such that a parallel boundary layer model can be applied.  Consequently, the irrotational velocity (free-stream velocity hereafter) in this model depends only on time
\begin{equation}
u_0^*(t^*)=U_{0m}^*\mathrm{sech}^2(-\omega^*t^*),
\label{eq:velAmb}
\end{equation}
where $U_{0m}^*$ is the maximum free-stream velocity and $\omega^*$ is the effective wave frequency. The reader is referred to \cite{onder_liu_2020} for the relation of these quantities to wave parameters. Using the wave frequency and kinematic viscosity, the Stokes length is defined
\begin{equation}
\delta_s^{*}=\sqrt{{2\nu^{*}}/{\omega^{*}}}
\label{eq:Stokes}
\end{equation}
as the length scale of the boundary layer and employed in the definition of Reynolds number 
\begin{equation}
\Rey_\delta=\frac{U_{0m}^*\delta_s^{*}}{\nu^{*}}.
\label{eq:Re}
\end{equation}
  The problem is defined in a Cartesian coordinate system $\vec x^*=(x^*,y^*,z^*)$, where $x^*$ is the direction of wave propagation (also called streamwise direction), $y^*$ is the spanwise direction parallel to wave crest, and $z^*$ is the vertical direction extending from the bed upwards. The velocity components associated with these directions are $\vec u^*=(u^*, v^*,w^*)$. We introduce the following normalizations to velocity fields, spatial coordinates, time and pressure, respectively: 
  \begin{equation}
  \vec u = \vec u^*/U_{0m}^*;~~\vec x=\vec x^*/\delta_s; ~~t=t^*\omega^*; ~~p=p^*/\rho^*U_{0m}^{*2}.  	
  \end{equation}
 The non-dimensional pressure gradient satisfying the free-stream momentum balance is given by
\begin{equation}
-\frac{\partial p_0}{\partial x}=\frac{4}{\Rey_\delta}\mathrm{sech}^2(-t)\mathrm{tanh}(-t).
\label{eq:dpdx}
\end{equation}
This pressure gradient drives the incompressible Navier-Stokes equations, which, together with the continuity equation, represent the governing equations for the three-dimensional instantaneous velocity in the boundary layer:
\begin{align}
\label{eq:mom}
	\frac{2}{\Rey_\delta}\frac{\partial u_i}{\partial t}+u_j \frac{\partial u_i}{\partial x_j}&=\frac{1}{\Rey_\delta} \frac{\partial^2 u_i}{\partial x_j\partial x_j}-\frac{\partial p}{\partial x_i}-\frac{\partial p_0}{\partial x}\delta_{i1}\\
	\label{eq:con}
	\frac{\partial u_i}{\partial x_i}&=0,
\end{align}
where summation over repeated indices are applied, and subscripts correspond to $(u_1,u_2,u_3)=(u,v,w)$ and $(x_1,x_2,x_3)=(x,y,z)$.

The bottom topography is composed of irregular corrugations, which are modelled by superposing two-dimensional sinosoidal modes equipped with a random amplitude $A_{nm}$ and a random phase $\phi_{nm}\in[0,2\pi]$, i.e., 
\begin{equation}
\eta(x,y)=h \sum_{n=0}^{L_x/r_x}\sum_{m=-L_y/r_y}^{L_y/r_y} A_{nm} \cos\left (\frac{2\pi n  x}{L_x}+\frac{2\pi m  y}{L_y}+\phi_{nm}\right),
\label{eq:eta}
\end{equation}
where $h$ is the roughness height, $r_x$ and $r_y$ are the cut-off wavelengths, and $L_x$ and $L_y$ are the length of the domain in the streamwise and spanwise directions, respectively. We have normalized the amplitudes of the sinusoidal components such that their Euclidean norm is unity, i.e, $\| A_{nm}\|=1$ and the mean mode is set to zero ($A_{00}=0$). A similar bottom model was applied in \cite{vittori1998direct} to represent bottom imperfections in an oscillatory boundary layer. \citeauthor{vittori1998direct} considered very small $h$, and modelled the wall using a Neumann boundary condition derived from the first-order Taylor expansion. In the present work, $h$ is not restricted to very small values and the corrugations are fully represented using coordinate transformation, cf. \S~\ref{sec:numerics} for details. 

When $h=0$, the boundary layer flow is one-dimensional and the laminar velocity field in figure~\ref{fig:uMean}c is obtained. When $h\neq 0$, the velocity field becomes three-dimensional due to perturbations introduced by bottom topography. As the bottom topography is generated using a different random set for $A_{nm}$ and $\phi_{nm}$ at each realization, ensemble averaging is required to analyze the flow fields. Ensemble averaging is defined as averaging over a dataset produced with the same $r_x$ and $r_y$.  Furthermore, spatial averaging over homogenous directions can be additionally imposed to accelerate the convergence of statistics, e.g., for the velocity field ensemble averaging is calculated by
\begin{equation}
\langle  {\vec u}  \rangle(z,t)=\frac{1}{RL_xL_y}\sum_{r=1}^{R}\int_0^{L_x}\int_0^{L_y}  {\vec u}^{\{r\}}(x,y,z,t)  \mathrm d x \mathrm d y,
\label{eq:planeAv}
\end{equation}
 where the velocities $\vec u^{\{r\}}$ ($r=1,\cdots ,R$) build a set of independent realizations. For simplicity, we will drop the supercript $\{r\}$ while referring to instantaneous fields hereafter. We denote the instantaneous fluctuating fields as $\vec u^{\prime}={\vec u}-\langle  {\vec u}  \rangle$. The highest level of the bed is denoted as $z_c$, and  the spatial averaging is only defined above this level.

\subsection{Numerical details}\label{sec:numerics}
The incompressible Navier--Stokes and continuity  equations in  (\ref{eq:mom}) and (\ref{eq:con}) are solved using the high-order spectral/hp element library Nektar++ \citep{cantwell2015nektar++}. Using the formulation in  \cite{Serson:2016ei}, the equations are first transformed to generalized coordinates $(\overline x,\overline y,\overline z)$   via the  
\begin{equation}
 x=\overline x,~~~~~ y=\overline y, ~~~~ z=\overline z+\mathrm {sech}^2(-\frac{\overline z}{L_m})\eta(x,y),
\end{equation}
with $L_m$ varying between $0.5-1$ depending on the case. This mapping transforms the physical domain with undulated bottom to a rectangular box, which is suitable for a mixed discretization, where a bi-dimensional spectral-element discretization \citep{karniadakis2005spectral} can be combined with Fourier expansions \citep{karniadakis1990spectral}.  The mixed representation allows  significant cost reduction and was employed in previous DNS works on bottom boundary layers \citep{onder_yuan_2019,onder_liu_2020,xiong2020bypass}.  We employ a bi-dimensional modified Legendre basis \citep{karniadakis2005spectral} in streamwise-wall normal ($\overline x-\overline z$) plane, and  Fourier expansions are defined in the spanwise ($\overline y$) direction. The $2/3$ rule is applied to avoid aliasing errors \citep{boyd2001chebyshev}. 


\begin{table}
\begin{center}
\begin{tabular}{ c  c c cc c c c    } 
\hline
Cases &  $h$ &  $R$& $z_c$&$(N_x^0,N_x^1,N_x^2)$ &$N_y$&$ (N_z^0,N_z^1,N_z^2)$ &$(N_p^0,N_p^1,N_p^2$)  \\[0.5ex]
\hline
h0.01 &  0.01   & 1 &0.027& (400,400,240)&320&(50,190,96) &(4,4,2)  \\
h0.04 &  0.04   & 1 &0.11&(480,480,320)&400&(60,228,128) &(5,5,3)  \\
h0.05 &  0.05   & 1 &0.137&(480,560,320)&480&(60,266,128) &(5,6,3)  \\
h0.055 &  0.055   & 1 &0.154&(480,560,320)&480&(60,266,128) &(5,6,3)  \\
h0.06 &  0.06   & 8 &0.165 &(560,560,320)&480&(70,266,128) &(6,6,3)  \\
h0.07 &  0.07 &1&0.192&(560,640,320)&720&(70,304,128) &(6,7,3)  \\


 \end{tabular} 
\end{center} 
\caption{Summary of cases. $h$ is the roughness height. Three subdomains defined in the vertical direction (see text for details). ($N_x^0$, $N_x^1$, $N_x^2$) and $(N_z^0,N_z^1,N_z^2)$ are the number of grid points in each subdomain in horizontal and vertical directions, respectively. $N_p$ is the polynomial orders of high-order finite elements. $R$ is the number of realizations. $R=8$ in case h0.06 is only applicable until the onset of transition ($t<\pi/9$). Beyond this time instant, only one realization is continued to study the breakdown $(R=1)$. $z_c$ is the highest elevation for the random bed distribution in figure~\ref{fig:wavywall}.  } 
\label{tab:cases}
\end{table}

\begin{figure}
\begin{center}

\includegraphics{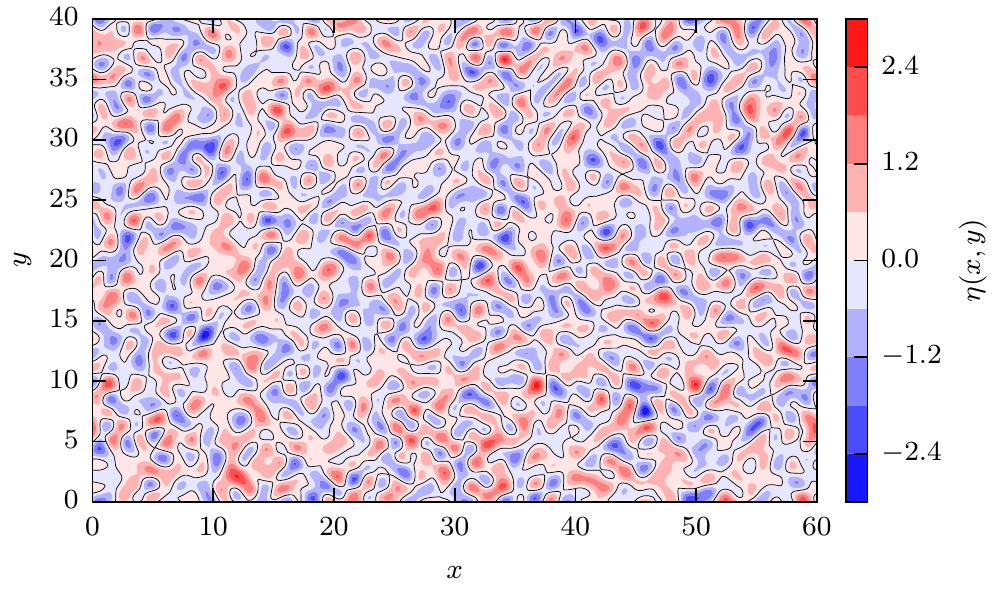}
\caption{\label{fig:wavywall} A randomly distributed bed topography with the roughness height $h=1$, and cut-off corrugation wavelengths $r_x=2$ and $r_y=2$ in streamwise and spanwise directions respectively, (\ref{eq:eta}).}
\end{center}
\end{figure}

The details of runs are given in table~\ref{tab:cases}.  For the present flow configuration, the transition scenario depends on four parameters: $\Rey_\delta$, $h$, $r_x$ and $r_z$. In this study, we investigate the receptivity and transition by varying the roughness height between $h=0.01$ and $0.07$. In order to fully resolve the roughness sublayer without excessive computational demand, the smallest corrugation wavelengths are set to a moderate value: $r_x=r_z=2$. An instance of this bed topography is shown in figure~\ref{fig:wavywall}. 
The Reynolds number is set to $\Rey_\delta=2000$. This is the highest Reynolds number in \cite{Sumer:2010ce}, where turbulent spots mediated the transition to turbulence in the corresponding case. We will show that a rich variety of transition scenarios are possible at this Reynolds number depending on the roughness height. 

The height, width and length of the computational domain are $L_x=60$, $L_y=40$ and $L_z=60$ (normalized with $\delta_s^*$), respectively.  The computational domain is sufficiently large to allow the transition modes and their secondary instabilities to develop, e.g., the vortex rollers have a streamwise spacing of about $\Delta x \approx 15$ \citep{Vittori:2008gv} and the linear non-normal theory predicts a spanwise streak spacing $\Delta y \approx 4-5$ \citep{Verschaeve:2017fa,onder_liu_2020}.  Periodic boundaries are employed in streamwise and spanwise directions.  The no-slip boundary condition ($\vec u=\vec 0$) is applied on the bottom wall and the zero-Neumann condition ($\partial \vec u/\partial z=\vec 0$) is applied at the top boundary.

The computational domain is discretized using a structured grid. In the vertical direction, the domain is partitioned into three subdomains: $\Omega_0:=\overline z \in [0,0.2]$, $\Omega_1:=\overline z \in[0.2,8]$, $\Omega_2:=\overline z \in [8,60]$.  $\Omega_0$ is designed to resolve the roughness sublayer with a finer resolution in the vertical with 10 elements, whose size increases gradually with an expansion ratio of 1.08 between adjacent elements. $\Omega_1$ is the domain where the transition and consequent turbulence takes place. This partition is designated with 38 elements in the vertical, and an expansion ratio of 1.05 is employed. In the outer most partition, $\Omega_2$,  32 elements are defined with an expansion ratio of 1.05. 80 elements are designated to the streamwise direction. The laminar flow in all cases are simulated using the polynomial order $N_p=4$ in $\Omega_0$ and $\Omega_1$,  $N_p=2$ in $\Omega_2$, and $60$ Fourier modes ($N_y=120$ spanwise grid points). Wall units are defined using the average skin-friction drag imposed by the bed, i.e., $\tau_b^*:=\sum_{\{r\}}(\int_\Gamma \vec F^{\nu,\{r\}} \mathrm d \Gamma)\cdot \hat{\vec e}_x/L_xL_yR$, where $\Gamma$ is the bottom surface, and $\vec F^{\nu,\{r\}}$ is the  viscous drag force at a point on the surface in a given realization. In a laminar SWBL over a flat bottom ($h=0$), the maximum mean skin-friction drag over the entire phase space is $\tau_b^{\max *}:=\max\{\tau_b^{*}\}=1.21\rho^*U_{0m}^{*2}\Rey_\delta$ (figure~\ref{fig:drag} in \S\ref{sec:breakdown}). This value is not significantly modified over the cases with rough topography, and can be employed for these cases also. At $\Rey_\delta=2000$, the maximum laminar friction velocity is $u_\tau^*:=\sqrt{\tau_b^{\max *}/\rho^*}=0.025U_{0m}^*$, and the corresponding viscous length scale is $\delta_{\nu}^*:=\nu^*/u_\tau^*=0.02\delta_s^*$. Consequently, in $\Omega_0$ and $\Omega_1$, the grid spacings in wall units (i.e. normalized with $\delta_{\nu}^*$) are $\Delta \overline x^+=7.38$ and $\Delta \overline y^+=16.4$ in homogenous directions. A spectral analysis in \S~\ref{sec:streaks} assesses the resolution in these directions for the receptivity stage and shows no spurious accumulation of energy in high wavenumbers.  In the vertical direction, the resolution at the wall is $\Delta \overline z^+=0.16$. As we aim to resolve all scales of the flow in our DNS experiments, no artificial stabilization technique such as spectral vanishing viscosity \citep{Kirby:2006hr} is employed. This results in significant numerical instabilities when the flows start to break down into finer scales during the transition. To avoid these numerical problems due to underresolution, we employed a p-type refinement \citep{karniadakis2005spectral}, in which polynomial orders and dimension of the Fourier space are increased until stability is achieved. These high-order refinements yield very dense grids, whose resolution details are presented in Table~\ref{tab:cases} for each case.  

 A system of differential algebraic equations is obtained using the continuous Galerkin method, and the coupled system is segregated using a velocity-correction scheme designed for transformed coordinates, cf. \cite{Serson:2016ei} for details and validations. The momentum equations are integrated in time using a second-order scheme, in which both advection and diffusion terms are treated implicitly \citep{vos2011generic}.  The additional viscous and pressure terms due to  coordinate transformation are solved explicitly whenever possible to benefit from the lower cost of the explicit scheme. In cases with high roughness (h0.06 and h0.07 in table~\ref{tab:cases}), the explicit scheme was unstable and we have switched to implicit mapping. A varying time-step size is utilized, which is refined adaptively with increasing velocities in the wave event. The maximum Courant--Friedrichs--Lewy (CFL) numbers in all cases are kept below $0.15$ to ensure a good temporal resolution.  All the resulting computational fields are mapped back to physical coordinates ($\vec x$) when analysing the results.   
 


\section{Receptivity stage: development of streaks}\label{sec:streaks}

\begin{figure}
\begin{center}
\includegraphics[]{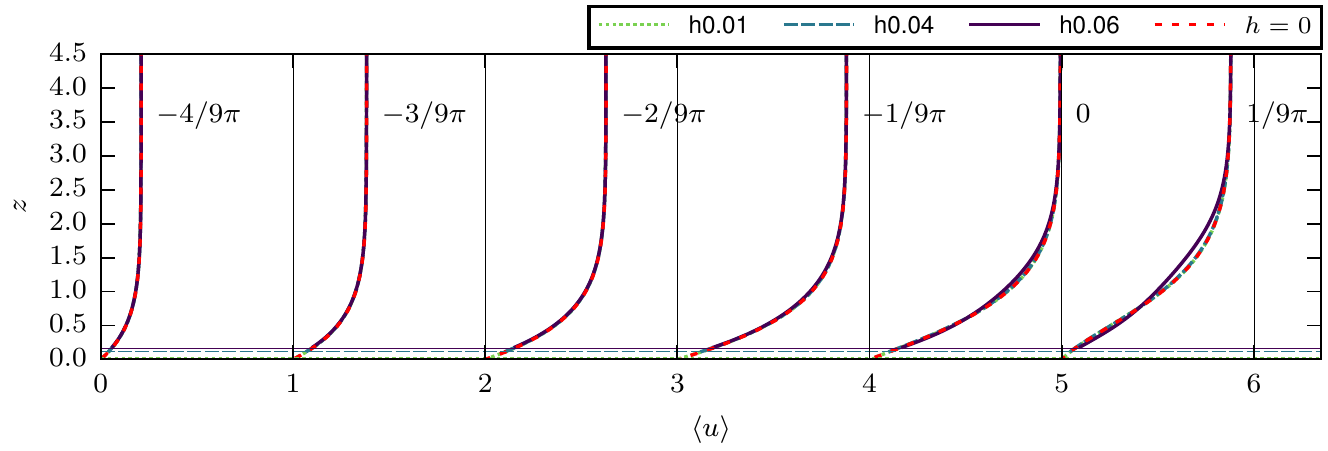}
\caption{ \label{fig:U} The vertical profiles of ensemble-averaged steamwise velocities between times $[-4/9\pi,1/9\pi]$ for cases~h0.01, h0.04 and h0.06. Reference profiles obtained over a perfectly smooth wall ($h=0$) are also plotted. The profiles are shifted by unity at each phase.The highest crest level ($z_c$) for each case is shown with an horizontal line of same type.}
\end{center}
\end{figure}

In this section, we study the response of the SWBL to bottom topography until the onset of transition. 
The vertical profiles of ensemble-averaged velocities $\langle u \rangle$ in cases h0.01, h0.04 and h0.06 are presented in figure~\ref{fig:U}. Velocity profiles over flat bottom ($h=0$) are also presented as reference.  The profiles over random topography start at $z=z_c$. Above this level, there is an excellent match among different cases until the end of FPG stage. Starting from phase $t=-1/9\pi$, noticeable deviations occur for h0.06, whereas cases~h0.01 and h0.04 appear to strictly follow the reference laminar profile for all presented times.
 
\begin{figure}
\begin{flushleft}
\pgfmathsetmacro{\xii}{3.80}
\pgfmathsetmacro{\dx}{6.0}
\pgfmathsetmacro{\xA}{-2.1}
\pgfmathsetmacro{\yA}{0.6}
\pgfmathsetmacro{\dyi}{-1.2}
\begin{tikzpicture}[declare function={swave(\x)=(1/(cosh(\x)^2);}]
\node (f1) at (\xii,0) {\includegraphics{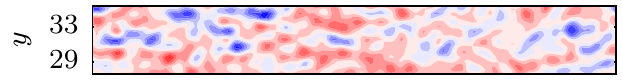} }+ (\xii+\xA+0.2,0+\yA) node[font=\small,color=black] {(a)}
+ (0,0) node[font=\scriptsize,color=black] {$-17/18\pi$}+(\xii+0.5,0.66) node[font=\large,color=black] {$u/u_0(t)$}
+(0.1,0.66) node[font=\large,color=black] {$t$};
\node (f1) at (\xii+\dx,0) {\includegraphics{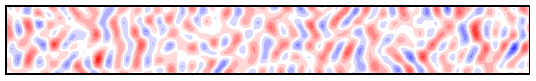} }+ (\xii+\dx+\xA-0.2,0+\yA) node[font=\small,color=black] {(b)}
+(\xii+\dx,0.66) node[font=\large,color=black] {$w/u_0(t)$};

\pgfmathsetmacro{\dy}{\dyi*1}
\node (f1) at (\xii,\dy) {\includegraphics{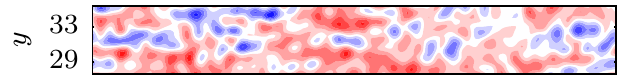} }+ (\xii+\xA+0.2,0+\yA+\dy) node[font=\small,color=black] {(c)}
+ (0,\dy) node[font=\small,color=black] {$-13/18\pi$};
\node (f1) at (\xii+\dx,+\dy) {\includegraphics{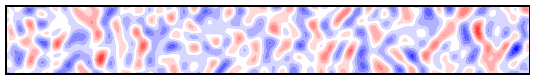} }+ (\xii+\dx+\xA-0.2,0+\yA+\dy) node[font=\small,color=black] {(d)};

\pgfmathsetmacro{\dy}{\dyi*2}
\node (f1) at (\xii,\dy) {\includegraphics{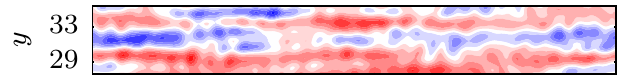} }+ (\xii+\xA+0.2,0+\yA+\dy) node[font=\small,color=black] {(e)}
+ (0,\dy) node[font=\small,color=black] {$-9/18\pi$};
\node (f1) at (\xii+\dx,+\dy) {\includegraphics{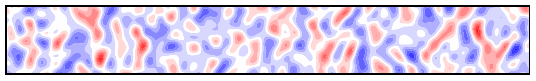} }+ (\xii+\dx+\xA-0.2,0+\yA+\dy) node[font=\small,color=black] {(f)};

\pgfmathsetmacro{\dy}{\dyi*3}
\node (f1) at (\xii,\dy) {\includegraphics{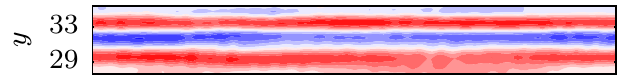} }+ (\xii+\xA+0.2,0+\yA+\dy) node[font=\small,color=black] {(g)}
+ (0,\dy) node[font=\small,color=black] {$-5/18\pi$};
\node (f1) at (\xii+\dx,+\dy) {\includegraphics{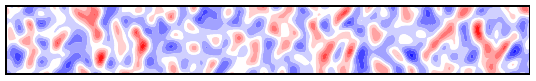} }+ (\xii+\dx+\xA-0.2,0+\yA+\dy) node[font=\small,color=black] {(h)};

\pgfmathsetmacro{\dy}{\dyi*4}
\node (f1) at (\xii,\dy) {\includegraphics{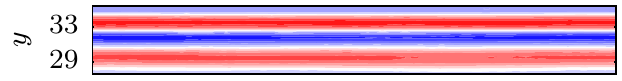} }+ (\xii+\xA+0.2,0+\yA+\dy) node[font=\small,color=black] {(i)}
+ (0,\dy) node[font=\small,color=black] {$-1/18\pi$};
\node (f1) at (\xii+\dx,+\dy) {\includegraphics{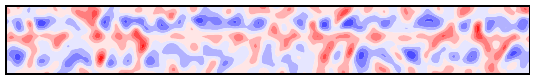} }+ (\xii+\dx+\xA-0.2,0+\yA+\dy) node[font=\small,color=black] {(j)};

\pgfmathsetmacro{\xio}{\xii}
\pgfmathsetmacro{\xii}{\xio+0.066}
\pgfmathsetmacro{\dy}{\dyi*5}
\pgfmathsetmacro{\dya}{-0.4}
\node (f1) at (\xii,\dy+\dya) {\includegraphics{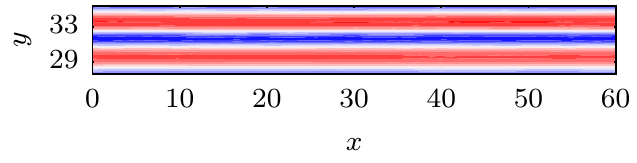} }+ (\xio+\xA+0.2,0+\yA+\dy) node[font=\small,color=black] {(k)}
+ (0.1,\dy) node[font=\small,color=black] {$3/18\pi$};
\node (f1) at (\xii+\dx,+\dy+\dya) {\includegraphics{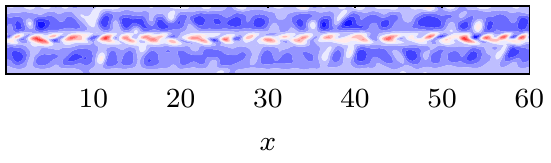} }+ (\xio+\dx+\xA-0.2,0+\yA+\dy) node[font=\small,color=black] {(l)};

\end{tikzpicture}
\caption{\label{fig:uw} Temporal evolution of streamwise ($a,c,e,g,i,k$) and vertical ($b,d,f,h,j,l$) velocities on plane $z=0.5$ between $28<y<34$. The velocities are normalized by the free-stream velocity $u_0(t)$ at the respective phase. The contours levels span 12 levels between: (a) [0.25,0.28]; (b) [-0.06,0.052]; (c) [0.52,0.6]; (d) [-0.034,0.037]; (e) [0.57,0.63]; (f) [-0.016, 0.02]; (g) [0.52,0.63]; (h) [-0.0083, 0.012]; (i) [0.26,0.69]; (j) [-0.0062,0.0073]; (k) [-0.04,0.53]; (l) [-0.0043, 0.008].}
\end{flushleft}
\end{figure}

\begin{figure}
\begin{flushleft}
\pgfmathsetmacro{\xii}{4.0}
\pgfmathsetmacro{\dx}{6.29}
\pgfmathsetmacro{\xA}{-2.1}
\begin{tikzpicture}[]

\node (f1) at (\xii,0) {\includegraphics[width=.71\textwidth]{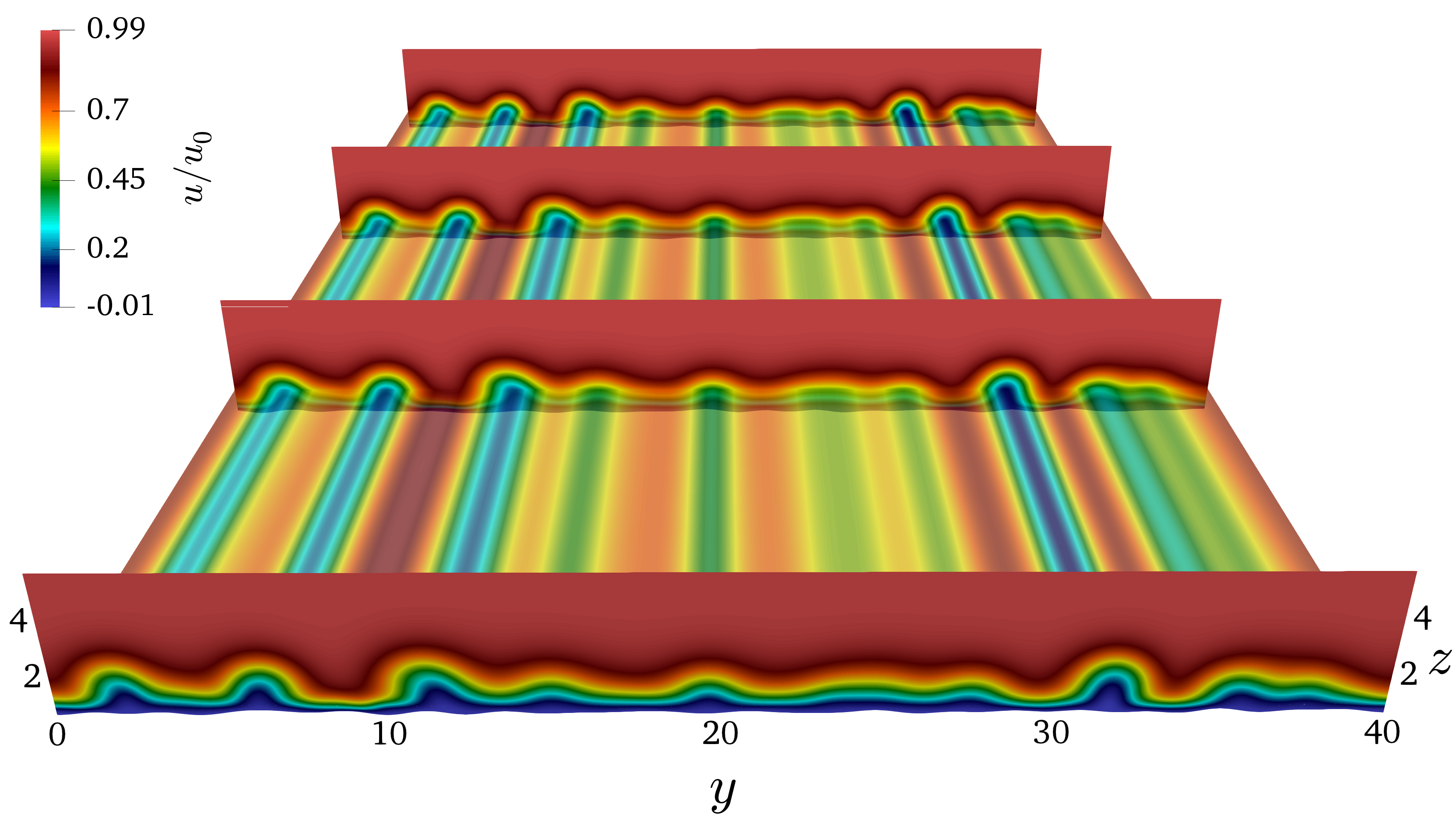} }+ (1.,2.5) node[font=\normalsize,color=black] {(a)};

\node (f2) at (\xii+\dx,1) {\includegraphics{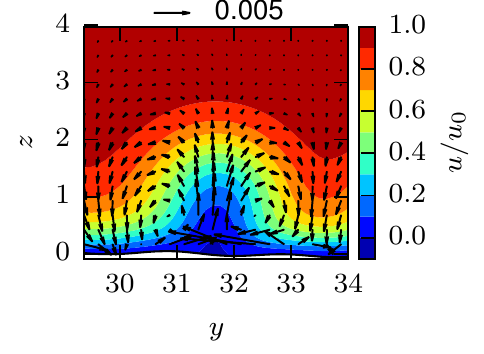} }+ (7.5,2.5) node[font=\normalsize,color=black] {(b)};

\draw[ultra thick] (6,-2.3) rectangle (7.1,-1.1);
\draw[ultra thick] (7.1,-2.2)--(10.04,-2.2) ;
\draw[-latex,ultra thick] (10.04,-2.2) -- (10.04,-0.75);

\draw[-latex, thick] (-0.5,-2)--(-0.84,-2.7);
\node[font=\normalsize,color=black] at (-0.5,-2.6)  {$x$};

\end{tikzpicture}
\caption{\label{fig:streak3D} The instantaneous fields at $t=1/9\pi$ in case h0.06. ($a$) Contours of streamwise velocity on a horizontal plane $z=0.5$ ($50\%$ opaque), and on four vertical cutplanes evenly spaced in the streamwise direction. The contours are normalized with the instantaneous free-stream velocity $u_0(t=1/9\pi)$.  ($b$) Enlarged view of a streak at cutplane $x=0$. Arrows show in-plane velocity components $(v \hat{\vec e}_y+ w \hat{\vec e}_z)/u_0$. }
\end{flushleft}
\end{figure}
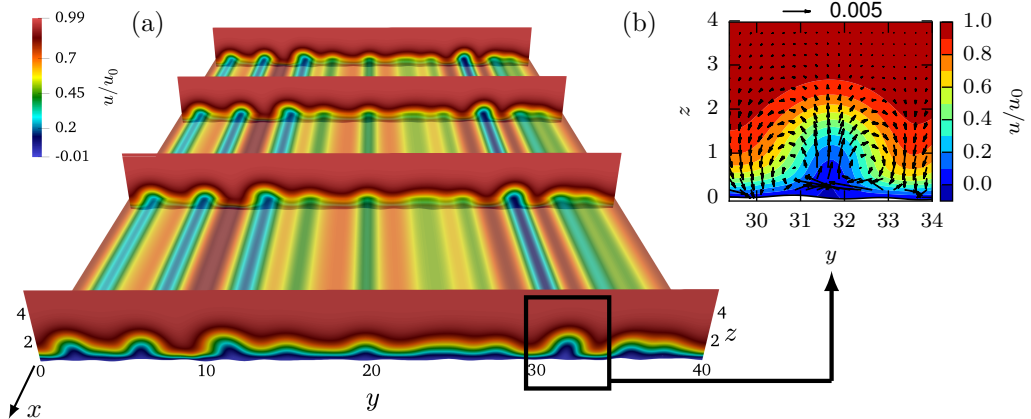

The evolution of streamwise and vertical instantaneous velocities on a subplane at $z=0.5$ are shown in figure~\ref{fig:uw} for case h0.06. In the initial stages of the event, both velocity components directly respond to bottom roughness and have irregular fluctuations on the plane (figures~\ref{fig:uw}$a$--$d$).  When the the presence of the wave becomes stronger, the boundary layer starts to amplify streaks of low and high streamwise momentum, cf. figures~\ref{fig:uw}($e,g$). At the end of the FPG stage, these structures already dominate the flow, and the streamwise velocity becomes approximately streamwise-constant. This can be also seen in figure~\ref{fig:streak3D}($a$), where $u(t=1/9\pi)$ is demonstrated in the full domain. The boundary layer is modulated along its span by streamwise-constant streaks. The prevelance of streamwise-constant streaks are consistent with the predictions of linear non-normal growth theory \citep{Verschaeve:2017fa,onder_liu_2020}. These streaks are accompanied by counter-rotating vortices (figure~\ref{fig:streak3D}$b$), which are arranged to transport low-momentum fluid upwards and high-momentum fluid downwards. This is the lift-up mechanism \citep{landahlJFM80}, which is discussed in detail in \cite{onder_liu_2020} for SWBLs. As the cross-stream velocity components building the vortices are of the same order (figure~\ref{fig:streak3D}$b$), we will consider hereafter only the vertical velocity to discuss the dynamics of vortices. The streamwise alignment of vertical velocity in figures~\ref{fig:uw}($j,l$) implies that the counter-rotating vortices are, like the streaks they generate, longitudinal  structures. However, unlike the streamwise streaks streamwise vortices do not dominate the momentum in their direction, as we observe considerable smaller-scale fluctuations in figures~\ref{fig:uw}($j,l$). This is because streamwise vortices are much weaker structures than the streamwise streaks as indicated by the values in the quiver key and colorbar in figure~\ref{fig:streak3D}($b$).  In fact, the linear streamwise-constant perturbation equations for smooth wall-bounded flows suggest that the cross-stream components are one order lower in Reynolds number than the streamwise components \citep{waleffe1995hydrodynamic,onder_liu_2020}. The same principle carries over here to the SWBL over rough bottom topography, and the amplification concentrates almost entirely in one (streamwise) component.

The development of longitudinal streaks and vortices can be further inspected using longitudinal and transverse spectra at a selected height.  The longitudinal spectral density for fluctuating velocities is approximated on the discrete grid by
  \begin{equation}
	E_{x,ij}(k_x,z,t)\approx\frac{2}{\Delta k_x} \langle \hat{u}_i (k_x,y,z,t)  \hat{u}_j^*(k_x,y,z,t) \rangle, \mbox{ ~~~~for } k_x\geq0,
		 \label{eq:Ex}
\end{equation}
where $\hat{\vec u}$ are the Fourier modes that are associated with the streamwise wavenumber $k_x$, $\Delta k_x=2\pi/L_x$ is the wavenumber spacing, and $\langle \cdot \rangle$ represents the combination of ensemble averaging and spatial averaging over the spanwise direction.  The subscript $x$ demonstrates the direction of Fourier decomposition, whereas $i$ and $j$ denote the fluctuating velocity components under the inspection, e.g., $E_{x,uw}$ is the longitudinal cross-spectral density function for components $u^{\prime}$ and $w^{\prime}$.  Integrating over spectral densities delivers the Reynolds stress, $\langle u_i^\prime u_j^\prime\rangle=\int E_{x,ij}\mathrm d k_x$. Similarly, the transverse spectral density function is obtained by
  \begin{equation}
	E_{y,ij}(k_y,z,t)\approx\frac{2}{\Delta k_y} \langle \hat{u}_i (k_y,x,z,t)  \hat{u}_j^*(k_y,x,z,t) \rangle, \mbox{ ~~~~for } k_y\geq0,
	\label{eq:Ey}
\end{equation}
where $ \hat{\vec u}(k_y,x,z,t) $ are the spanwise Fourier modes associated with $k_y$, and $\langle \cdot \rangle$ represents this time  combination of the ensemble averaging and spatial averaging over the streamwise direction. 



\begin{figure}
\begin{center}
\includegraphics[]{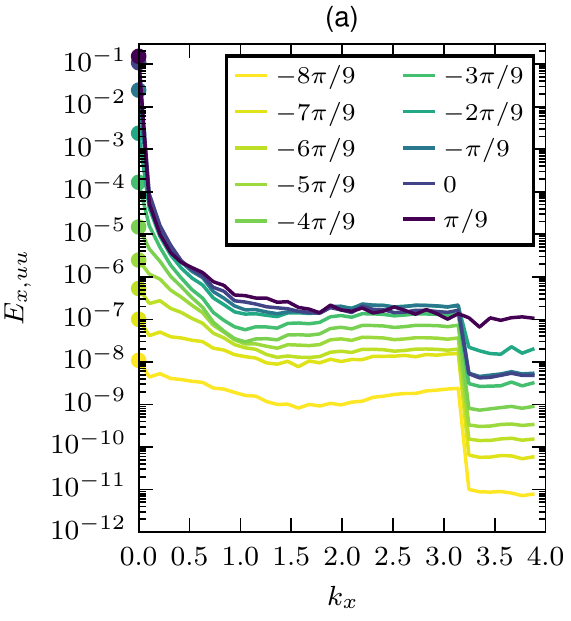}
~~~~~~~\includegraphics[]{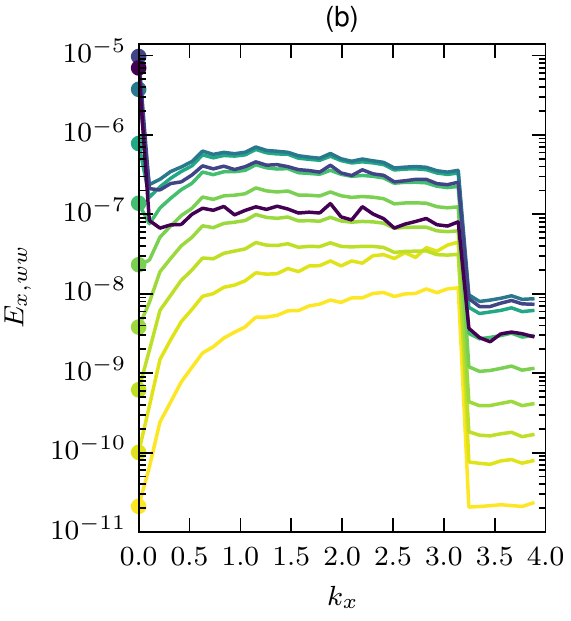}
\caption{ \label{fig:Ex} Longitudinal spectral densities (\ref{eq:Ex}) for streamwise ($a$) and vertical ($b$) velocity fluctuations at $z=0.5$ at phases $t=\{-8\pi/9,-7\pi/9,\cdots,\pi/9\}$ for case h0.06. The color coding is the same for both figures. The energy in streamwise-constant modes ($k_x=0$) are shown with circles. }
\end{center}
\end{figure}

Figure~\ref{fig:Ex} presents the evolution of the longitudinal spectral densities (\ref{eq:Ex}) for streamwise and vertical velocity fluctuations at $z=0.5$ for case h0.06. Wall undulations with streamwise cut-off wavenumber $r_x=2$ (cf. figure~\ref{fig:wavywall}) are defined in the spectral band $0\leq k_x\leq \pi$, i.e., (\ref{eq:eta}). Therefore, the flow is subject to a broadband forcing in this spectral band. In this regard, a step-like abrupt decay is observed at the end of both spectra at $k_x=\pi$ due to cut-off in forcing.  The boundary layer responds evenly to broadband forcing in $1\lesssim k_x<\pi$, and we observe a flat spectrum in this band. Below this band, the response is far greater due to intrinsic noise amplification in the boundary layer and  the streamwise-constant mode ($k_x=0$) prevails with a very distinctive peak in the streamwise velocity spectra (figure~\ref{fig:Ex}$a$). Streamwise-constant modes are initially negligible in vertical velocity spectra (figure~\ref{fig:Ex}$b$). They only become prevalent in the late FPG stage (starting from $t=-3\pi/9$ in the figure). Interestingly, this late amplification is limited to streamwise-constant modes while the rest of low-frequency band remains below the plateau level. This fine-tuned amplification hints to a mechanism, in which energetic streamwise-constant streaks induce streamwise-constant vortices. Details of this process will be investigated later in this section. Consistent with the visualizations in figure~\ref{fig:uw}, streamwise-constant  vertical motions are not as dominant as their streamwise counterparts, e.g., the peaks at $k_x=0$ at late phases are located 5-6 decades higher than the plateau $1\lesssim k_x<\pi$ in figure~\ref{fig:Ex}$a$, whereas the difference is only 1-2 decades in figure~\ref{fig:Ex}$b$. 

\begin{figure}
\begin{center}
\includegraphics[]{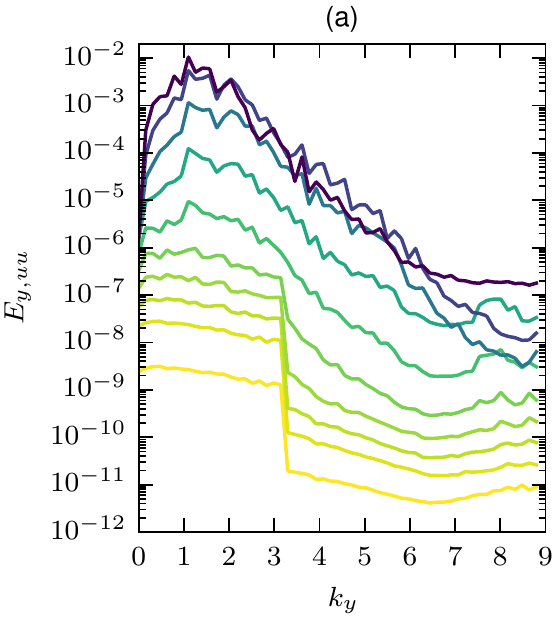}
~~~~~~~\includegraphics[]{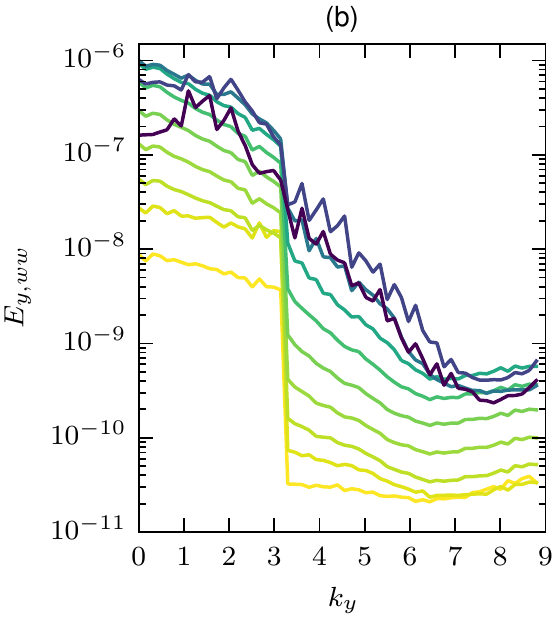}
\caption{ \label{fig:Ey} Transverse spectral densities (\ref{eq:Ey}) for streamwise ($a$) and vertical ($b$) velocity fluctuations at $z=0.5$ at phases $t=\{-8\pi/9,-7\pi/9,\cdots,\pi/9\}$ for case h0.06. See figure~\ref{fig:Ex}a for the color coding of the lines.}
\end{center}
\end{figure}

Transverse velocity spectra at $z=0.5$ are presented in figure~\ref{fig:Ey} for case h0.06. The forcing by bottom topography focuses in the band $0\leq k_y\leq \pi$ as the spanwise cut-off wavenumber of the bed undulations is $r_y=2$. At earlier times we see again a step-like decay for $k_y>\pi$. However, in the transverse spectra, the energy spreads soon to higher wavenumbers and the step profile disappears. This is due to the development of internal shear layers around streaks, which promotes fine-scale energy.  A peak at $k_y\approx 1.5$ starts to appear in the streamwise energy spectra  at $t\approx-\pi/3$ and becomes more prevalent at later times, cf figure~\ref{fig:Ey}($a$). These peaks represent the average spanwise spacing between streaks.  Using a linear non-normal analysis based on a body forcing model, \cite{onder_liu_2020}  also observed a high amplification in the range $k_y\approx1.5$. This suggests that a similar non-normal amplification mechanism (lift-up mechanism) becomes prominent for $t\geq-\pi/3$ in the present problem. A peak is observed only at $t=\pi/9$ in the vertical spectra, cf figure~\ref{fig:Ey}$b$. Besides the relative weakness of streamwise-constant vertical fluctuations compared to background fluctuations,  this delayed  appearance of the peak in the vertical spectra also suggests again a mechanism, in which streamwise-constant streaks, when they become sufficiently strong, feed streamwise-constant vortices. 


The receptivity process can be further elaborated by studying the dynamics of streamwise streaks and vortices in isolation.  \cite{onder_meyers_2018} employed a simple sharp spectral filter to distinguish between long streaks in the atmospheric boundary layer and wakes of wind turbines. It is based on filtering out streamwise wavenumbers above a cut-off wavenumber $k_x^c$ , i.e., in Fourier space:
 \begin{equation}
     \mathrm G(k_x;k_x^c)=\left\{
                \begin{array}{ll}
                  1, \text{  if } \left |k_x\right |\leq k_x^c,\\
                  0, \text{  if } \left | k_x \right |> k_x^c.
                \end{array}
              \right.
              \label{eq:filter}
 \end{equation}
We adapt the same filter here to extract long streaky components, 
\begin{equation}
\vec  {\widetilde {u^\prime}}(\vec x,t):=\mathrm G \circ  {\vec u}^{\prime}(\vec x,t)
\label{eq:Gop}
\end{equation}
The residual finer scale velocity components are then expressed by $ \vec  {u}^\dprime:= \vec  {u}^{\prime}-\widetilde{\vec u^\prime}$, i.e., $\mathrm G \circ  {\vec u}^{\prime\prime}=\widetilde {\vec u^\dprime}= 0$. This results in the following triple decomposition of the instantaneous velocity field
\begin{equation}
\vec u= \langle \vec u \rangle +\underbrace{\vec  {\widetilde{u^\prime}} + \vec  {u}^\dprime}_{\vec {u}^{\prime}}.
\end{equation}
This decomposition is defined only in the fluid domain above the highest topography ($z>z_c$). We note that $\langle \widetilde {u^\prime_i}  u_j^\dprime \rangle=0$, as  $\widetilde {\vec u^\prime}=\sum_{k_1} \hat {\vec u} \ee^{\ii k_x x}$ and $\vec u^\dprime=\sum_{k_2} \hat {\vec u} \ee^{\ii k_x x}$ belong to separate spectral bands, i.e., the sets $k_1$ and $k_2$ have no common wavenumber, and Fourier modes corresponding to different wavenumbers are uncorrelated, i.e.,  $\langle \hat u_i(k_x=\alpha,y,z,t) \hat u_j^*(k_x=\beta,y,z,t) \rangle = 0$ unless $\alpha=\beta$, cf. Appendix E.2 in \cite{pope2000} for details. In this regard, Reynolds stresses can also be decomposed into two components
\begin{equation}
\langle  u_i^{\prime}  u_j^{\prime}\rangle=\langle \widetilde {u^\prime_i} \widetilde {u^\prime_j}\rangle+\langle  u_i^\dprime  u_j^\dprime\rangle.
\label{eq:reyVLSM}
\end{equation}
We have seen above that streamwise-constant motions clearly dominate. Therefore, $k_x^c=0$ is selected as the cut-off wavenumber. The resulting filter acts like a spatial averaging operator over the streamwise direction, and designates the filtered fields the following additional properties 
\begin{equation}
\frac{\partial \widetilde {\vec u^\prime}}{\partial x}=\vec 0;~ \widetilde {\widetilde {\vec u^\prime} \vec u^{\prime\prime}}=\vec 0; ~ \widetilde {\widetilde {\vec u^\prime} \widetilde{\vec u^\prime}}=\widetilde {\vec u^\prime} \widetilde{\vec u^\prime}.
\label{eq:filProp}
\end{equation}




\begin{figure}
\begin{center}
\includegraphics[]{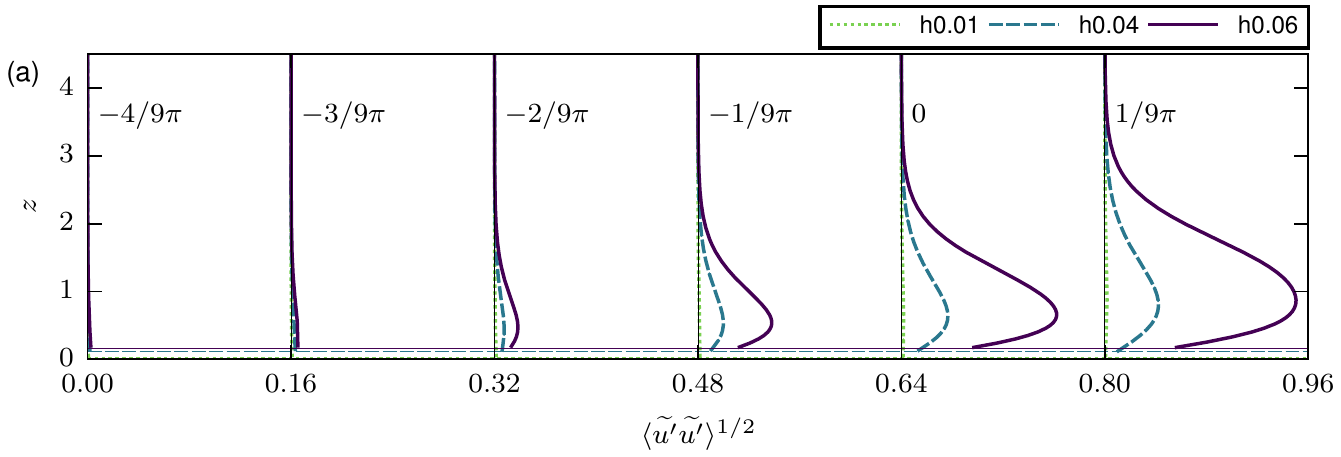}
\includegraphics[]{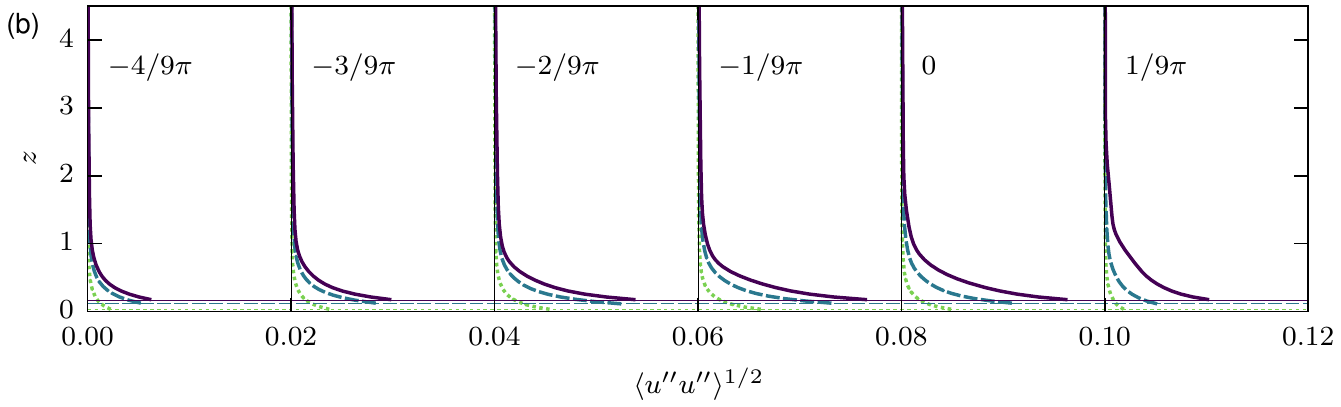}
\caption{ \label{fig:uu} Vertical profiles of streamwise velocity fluctuations for Cases h0.01, h0.04 and h0.06. ($a$) Intensity of streamwise-constant fluctuations $\langle \widetilde {u^\prime} \widetilde {u^\prime}\rangle^{1/2}$. The profiles are shifted by 0.16 at each phase. ($b$) Intensity of three-dimensional fluctuations $\langle u^\dprime  u^\dprime \rangle^{1/2}$. The profiles are shifted by 0.02 at each phase. The highest crest level ($z_c$) for each case is shown with an horizontal line of the same type. }
\end{center}
\end{figure}

The vertical profiles of streamwise fluctuation intensities for streamwise-constant  ($\langle \widetilde {u^\prime}\widetilde {u^\prime}\rangle^{1/2}$) and residual ($\langle u^{\prime\prime}  u^{\prime\prime} \rangle^{1/2}$) motions in cases h0.01, h0.04 and h0.06 are plotted in figure~\ref{fig:uu}($a$) and \ref{fig:uu}($b$), respectively. In all three cases, residual fluctuations peak at $z=z_c$ and decay rapidly upwards from there. Their intensities peak at $t=-1/9\pi$ when the bottom shear is maximum (cf.~figure~\ref{fig:drag} in \S\ref{sec:breakdown}).  Streamwise-constant fluctuations have very different characteristics compared to residual fluctuations. The peaks are located significantly above the crest level and move progressively to higher levels with lifting up of streaks, cf figure~\ref{fig:uu}($a$). Their intensities are an order of magnitude or more higher than residual intensities. The peak values in $\langle \widetilde {u^\prime}\widetilde {u^\prime}\rangle^{1/2}$ profiles represent an average value for streak amplitudes. These amplitudes increase with roughness height $h$ with an increasingly nonlinear rate, i.e., their scaling with roughness height is $h^p$ with $p>1$.  There is a significant jump between cases, e.g., at $t=1/9\pi$, the peak intensities are $0.0018$, $0.042$ and $0.15$ for h0.01, h0.04 and h0.06, respectively. Case h0.01 develops very weak streaks whose intensity is almost indistinguishable in figure~\ref{fig:uu}($a$). This suggests a nonlinear threshold mechanism for onset of streak amplification.  

\begin{figure}
\begin{center}
\includegraphics[]{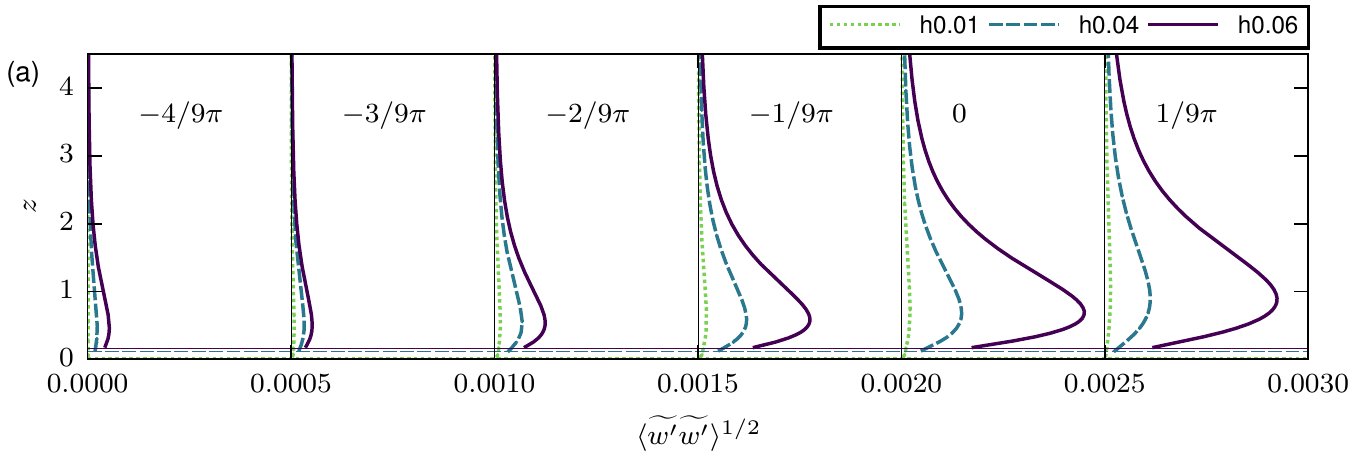}
\includegraphics[]{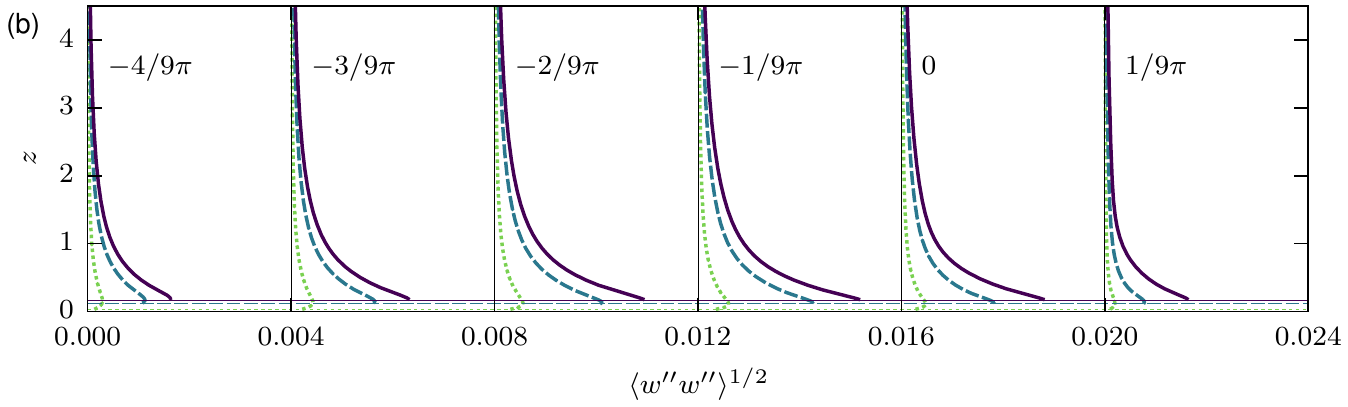}
\caption{ \label{fig:vv} Vertical profiles of vertical velocity fluctuations for Cases h0.01, h0.04 and h0.06. ($a$) Intensity of streamwise-constant fluctuations $\langle \widetilde {w^\prime} \widetilde {w^\prime}\rangle^{1/2}$. The profiles are shifted by 0.0005 at each phase. ($b$) Intensity of three-dimensional fluctuations $\langle w^\dprime  w^\dprime \rangle^{1/2}$. The profiles are shifted by 0.004 at each phase. The highest crest level ($z_c$) for each case is shown with an horizontal line of same type.}
\end{center}
\end{figure}

Figure~\ref{fig:vv}($a$) shows the profiles of streamwise-constant component of the vertical fluctuation intensities ($\langle \widetilde {w^\prime} \widetilde {w^\prime}\rangle^{1/2}$). The peak values of $\langle \widetilde {w^\prime} \widetilde {w^\prime}\rangle^{1/2}$ represents an average measure for the amplitude of streamwise-constant vortices. Similar to $\langle \widetilde {u^\prime} \widetilde {u^\prime}\rangle^{1/2}$, $\langle \widetilde {w^\prime} \widetilde {w^\prime}\rangle^{1/2}$ peaks significantly above the crest levels ($z_c$) and the relationship to the roughness height is nonlinear. As discussed above, these intensities are an order of magnitude lower in Reynolds number. At $t=1/9\pi$, the peak values are $5.95\times 10^{-6}$, $1.86\times 10^{-4}$ and $1.06\times 10^{-3}$ for h0.01, h0.04 and h0.06, respectively. These values are lower than the peak residual fluctuation intensities $\langle w^\dprime  w^\dprime \rangle^{1/2}$ in figure~\ref{fig:vv}b. 

\begin{figure}
\begin{center}
\includegraphics[]{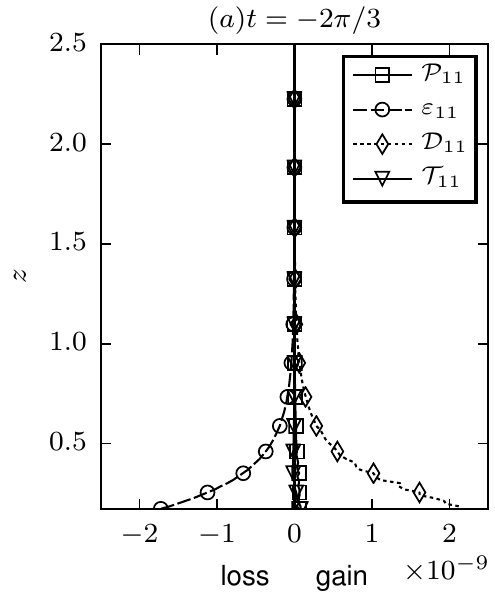}
\includegraphics[]{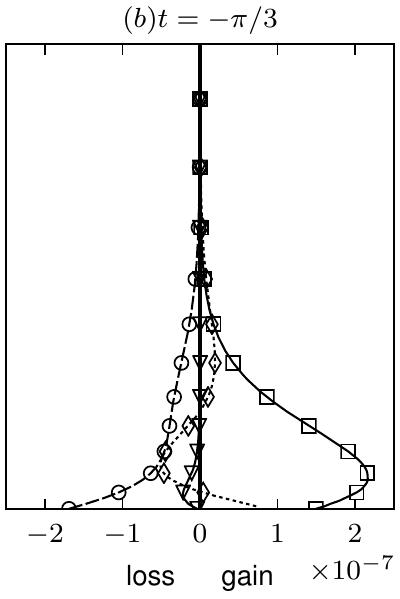}
\includegraphics[]{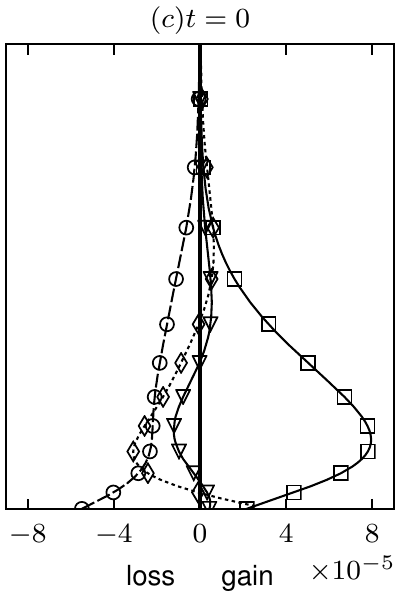}
\caption{ \label{fig:Euu} Energy budget for streamwise-constant streamwise fluctuations ($\widetilde {u^\prime}$),  (\ref{eq:Euu}), in case h0.06. ($a$) $t=-2\pi/3$; ($b$) $t=-\pi/3$; ($c$) $t=0$. }
\end{center}
\end{figure}

The relationship between bottom topography and the spatial organization of streaks, and the origins of nonlinear receptivity process remain to be elaborated. Some insights can be obtained by analysing the perturbation and energy equations. The governing equations for the ensemble-averaged velocity field $\langle \vec u \rangle=(\langle  u \rangle,0,0)$ in the region above the topography ($z>z_c$) are expressed by
\begin{align}
\label{eq:uMean}
	\frac{2}{\Rey_\delta}\frac{\partial \langle u \rangle}{\partial t}&=\frac{1}{\Rey_\delta} \frac{\partial^2 \langle u \rangle }{\partial z^2}-\frac{\partial \langle u^{\prime}w^{\prime}\rangle}{\partial z}- \frac{\partial p_0+\langle p \rangle}{\partial x},\\
	\label{eq:wMean}
	0&=-	\frac{\partial \langle w^{\prime}w^{\prime}\rangle}{\partial z}- \frac{\partial  \langle p \rangle}{\partial z}.
\end{align}
We obtain the governing equations for the fluctuating velocity fields  $\vec u^{\prime}$  by subtracting (\ref{eq:uMean})--(\ref{eq:wMean}) from (\ref{eq:mom}):
\begin{align}
\label{eq:uPMom}
		\frac{2}{\Rey_\delta}\frac{\partial u^{\prime}}{\partial t}+
		\langle u\rangle\frac{\partial u^{\prime}}{\partial x}+w^{\prime}\frac{\partial \langle u \rangle}{\partial z}
		+\nabla \cdot  (u^{\prime} \vec u^{\prime})
	&=
	\frac{1}{\Rey_\delta}\nabla^2 u^{\prime}
		+\frac{\partial \langle u^{\prime}w^{\prime}\rangle}{\partial z}-\frac{\partial p^{\prime}}{\partial x},\\
		\frac{2}{\Rey_\delta}\frac{\partial v^{\prime}}{\partial t}
			+\nabla \cdot  (v^{\prime} \vec u^{\prime})
	&=\frac{1}{\Rey_\delta}\nabla^2 v^{\prime}-\frac{\partial p^{\prime}}{\partial y},\\
	\label{eq:wPMom}
					\frac{2}{\Rey_\delta}\frac{\partial w^{\prime}}{\partial t}
	+\nabla \cdot  (w^{\prime} \vec u^{\prime})
	&=\frac{1}{\Rey_\delta}\nabla^2 w^{\prime}
	+\frac{\partial \langle w^{\prime}w^{\prime}\rangle}{\partial z}-\frac{\partial p^{\prime}}{\partial z},
\end{align}
where $\nabla= \partial /\partial x \hat {\vec e}_x+\partial /\partial y \hat {\vec e}_y+\partial /\partial z \hat {\vec e}_z$  and $\nabla^2=\nabla\cdot\nabla$. The governing equations for  streamwise-constant fluctuating fields $\widetilde {\vec u^\prime}$ are obtained by applying the filtering operation (\ref{eq:Gop}) to the individual terms in the fluctuation equations (\ref{eq:uPMom})--(\ref{eq:wPMom}) and imposing further simplications using (\ref{eq:filProp}):
\begin{align}
\label{eq:uCMom}
		\frac{2}{\Rey_\delta}\frac{\partial \widetilde {u^\prime}}{\partial t}
		+\widetilde {w^\prime}\frac{\partial \langle u \rangle}{\partial z}
		+\frac{\partial \widetilde {u^{\prime}v^{\prime}}}{\partial y}+\frac{\partial \widetilde {u^{\prime}w^{\prime}}}{\partial z}
	&=\frac{1}{\Rey_\delta}\widetilde \nabla^2 \widetilde {u^\prime}
	+\frac{\partial  \langle u^{\prime}w^{\prime} \rangle}{\partial z},\\
	\label{eq:vCMom}
		\frac{2}{\Rey_\delta}\frac{\partial \widetilde {v^\prime}}{\partial t}
		+\frac{\partial \widetilde {v^{\prime}v^{\prime}}}{\partial y}+\frac{\partial \widetilde {v^{\prime}w^{\prime}}}{\partial z}
	&=\frac{1}{\Rey_\delta}\widetilde \nabla^2 \widetilde {v^\prime}
	-\frac{\partial \widetilde {p^\prime}}{\partial y},\\
	\label{eq:wCMom}
		\frac{2}{\Rey_\delta}\frac{\partial \widetilde {w^\prime}}{\partial t}
		+\frac{\partial \widetilde {v^{\prime}w^{\prime}}}{\partial y}+\frac{\partial \widetilde {w^{\prime}w^{\prime}}}{\partial z}
	&=\frac{1}{\Rey_\delta}\widetilde \nabla^2 \widetilde {w^\prime}
	+\frac{\partial  \langle w^{\prime}w^{\prime} \rangle}{\partial z}
	-\frac{\partial \widetilde {p^\prime}}{\partial z},
	\end{align}
where $\widetilde \nabla= \partial /\partial y \hat {\vec e}_y+\partial /\partial z \hat {\vec e}_z$. These equations are supplemented with the pressure-Poisson equation for the filtered pressure, which is obtained by taking y-derivative of (\ref{eq:vCMom}) and z-derivative of  (\ref{eq:wCMom}) and then summing up the resulting equations:
	\begin{equation}
	\label{eq:PPE}
			-\left (\frac{\partial^2 \widetilde {p^\prime}}{\partial y^2}+\frac{\partial^2 \widetilde {p^\prime}}{\partial z^2}\right )
	=\frac{\partial^2 \widetilde {v^{\prime}v^{\prime}}}{\partial y^2}+2\frac{\partial^2  \widetilde {v^{\prime}w^{\prime}}}{\partial y\partial z}
+\frac{\partial^2  \widetilde {w^{\prime}w^{\prime}}}{\partial z^2}
-\frac{\partial^2   \langle w^{\prime}w^{\prime} \rangle}{\partial z^2}.
	\end{equation}
	Equation (\ref{eq:uCMom}) is the momentum equation for streaks. Equations (\ref{eq:vCMom})-(\ref{eq:PPE}) are the governing equations for streamwise-constant vortical motions. We note that streamwise-constant pressure is decoupled completely from streamwise fluctuations. It is observed that streak and vortex equations are only connected by nonlinear terms containing the residual fluctuations. We will see later this section that this link plays a key role for the feedback from streaks to vortices. Multiplying (\ref{eq:uCMom}) with $\widetilde {u^\prime}$ and ensemble-averaging the resulting equation, we obtain the energy budget for streaks
\begin{equation}
\begin{split}
\label{eq:Euu}
	\frac{1}{\Rey_\delta}\frac{\partial \langle \widetilde {u^\prime}^2 \rangle}{\partial t}
	&=\underbrace{- \left \langle \widetilde {u^\prime} \left (\frac{\partial \widetilde{  u^{\prime} v^{\prime}}}{\partial y}+\frac{\partial \widetilde{  u^{\prime} w^{\prime}}}{\partial z}\right)\right \rangle}_{\mathcal T_{11}}
	\underbrace{-\langle \widetilde {u^\prime} \widetilde {w^\prime}\rangle \frac{\partial \langle u \rangle}{\partial z}}_{\mathcal P_{11}}\\
	&+\underbrace{\frac{1}{2\Rey_\delta}\frac{\partial^2 \langle \widetilde {u^\prime} \widetilde {u^\prime} \rangle}{\partial z^2}}_{\mathcal D_{11}}
	\underbrace{-\frac{1}{\Rey_\delta}  \left \langle \left(\frac{ \partial \widetilde {u^\prime}}{\partial y}\right)^2+\left(\frac{ \partial \widetilde {u^\prime}}{\partial z}\right)^2\right \rangle}_{\varepsilon_{11}},
\end{split}
\end{equation}
where $\mathcal T_{11}$ contains the terms for mean and turbulent transport and redistribution, $\mathcal P_{11}$ is the production term, $\mathcal D_{11}$ is the diffusive transport term, and $\varepsilon_{11}$ is the dissipation term. Figure~\ref{fig:Euu} demonstrates the vertical profile of each budget term at three representative phases $t=-2\pi/3,-\pi/3$ and $0$ for case h0.06. Two different streak generation mechanisms are observed. Initially, when the free-stream velocity and boundary layer shear are still weak, the streaks are generated by diffusive transport from the bed upwards, cf.~figure~\ref{fig:Euu}($a$). At this initial stage, the production term is negligible, and vertical fluctuations and lift-up mechanism play no role.  At later phases of the FPG stage, vertical fluctuations and shear strengthen, and the lift-up mechanism is activated. Consequently, the  production term becomes the dominant gain term, cf.~figures~\ref{fig:Euu}($b$) and \ref{fig:Euu}($c$). 

\begin{figure}
\begin{center}
\includegraphics[]{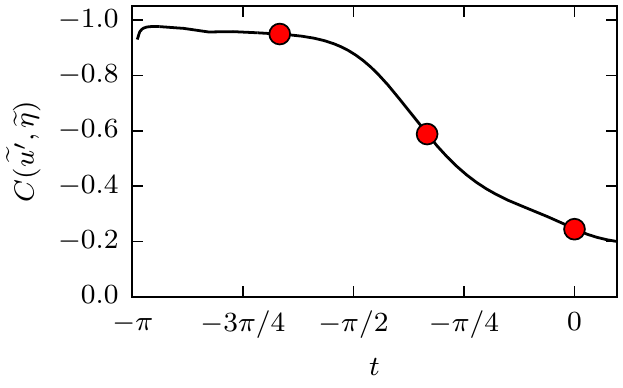}
\caption{ \label{fig:Cuy} Time evolution of the correlation coefficient $C(\widetilde {u^\prime}, \widetilde \eta)$, (\ref{eq:C}), between streamwise-constant bed elevation and streamwise-constant velocity $\widetilde {u^\prime}$ at $z=0.5$ for case h0.06. Red markers show the phases, for which $\widetilde {u^\prime}$ anf $\widetilde \eta$ are shown in figure~\ref{fig:uCy}.}
\end{center}
\end{figure}

\begin{figure}
\begin{center}
\includegraphics[]{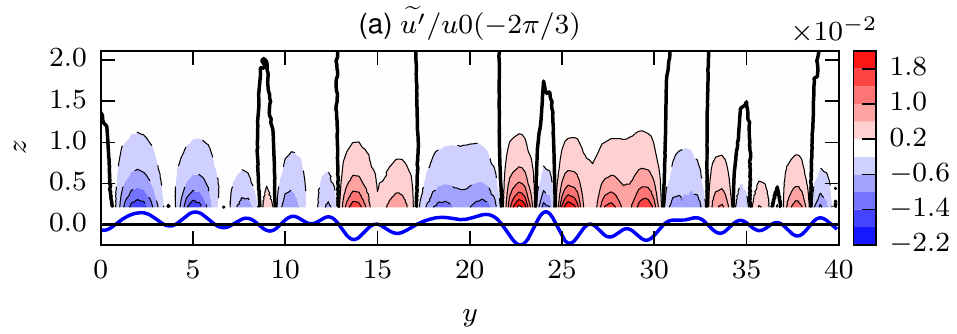}
\includegraphics[]{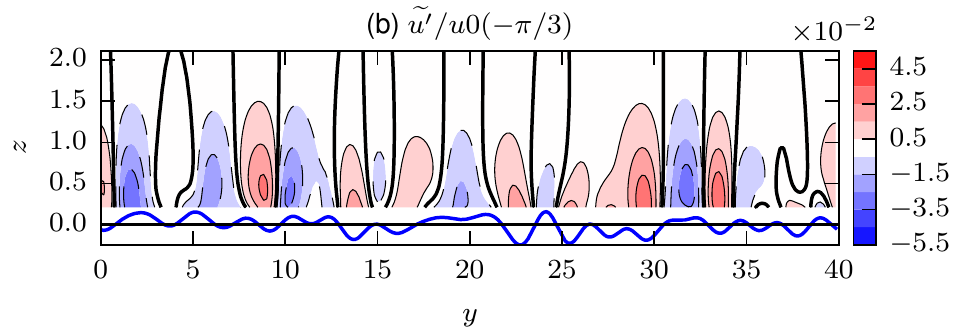}
\includegraphics[]{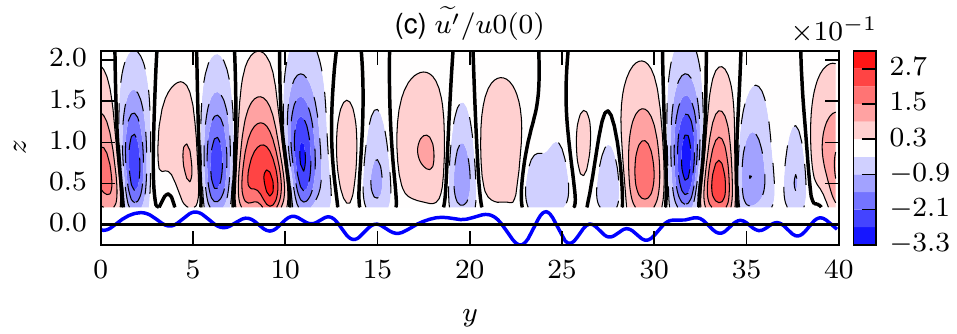}
\caption{ \label{fig:uCy} Contours streamwise-constant fluctuation velocity $\widetilde {u^\prime}$ are shown along with the streamwise-constant bed elevation $\widetilde \eta$ in case h0.06. ($a$) $t=-2\pi/3$; ($b$) $t=-\pi/3$; ($c$) $t=0$. Contours are normalized with the local free-stream velocity at the respective phases ($u_0$). The thick contour lines show the level $\widetilde {u^\prime}=0$. Bed elevation is magnified 12 times for visibility. }
\end{center}
\end{figure}

Although the energy of the early streaks are very low, they determine the initial positioning over the randomly distributed bed undulations. As the early streaks are produced by diffusive transport upwards from the bed,  direct connections between streak locations and bed topography are to be expected. Assuming a linear process, streamwise-constant streaks are excited by streamwise-constant modes of the topography. This relationship can be quantified using the correlation coefficient,  
\begin{equation}
\label{eq:C}
C(\widetilde {u^\prime},\widetilde \eta,z,t)=\frac{\langle \widetilde {u^\prime}(y,z,t) \widetilde \eta(y) \rangle}{\sqrt{\langle \widetilde {u^\prime}^2\rangle}\sqrt{ \langle \widetilde \eta^2\rangle}},
\end{equation}
where $\widetilde \eta=\mathrm G(k_x;k_x^c=0)\circ \eta$ is the filtered bed elevation function. The time evolution of $C(\widetilde {u^\prime},\widetilde \eta)$ at $z=0.5$ is plotted in figure~\ref{fig:Cuy} for case h0.06. Until $t\approx -\pi/2$, there is almost a perfectly negative correlation between the filtered topography and streaks, i.e., $C(\widetilde {u^\prime},\widetilde \eta)\approx -1$. Therefore, at this initial stage, low-speed streaks develop on high filtered topography and vice versa. The anticorrelations reduce in the second half of the FPG stage when the lift-up mechanism takes over the diffusive generation of streaks. However,  the energetic streaks at these later stages build on the orientation history before them. This is shown in figure~\ref{fig:uCy} for three representative time instances $t=-2\pi/3,-\pi/3$ and $0$ in case h0.06.  Initially, the relation between high topography and low-speed streaks and low topography and high-speed streaks is very strong as expected from cross-correlations, figure~\ref{fig:uCy}($a$). This association reduces with streaks getting stronger but it never completely disappears (figures~\ref{fig:uCy}$b,c$). In fact, the most unstable streak for this realization is the low-speed streak at $y=32$, which breaks down into turbulent spots at early APG stage (cf. \S~\ref{sec:breakdown}). We see that this streak is initially seeded by a wide bump in $30<y<33$ in the diffusive growth stage (figure~\ref{fig:uCy}$a$), and it grows further from there in the lift-up stage. 

\begin{figure}
\begin{center}
\includegraphics[]{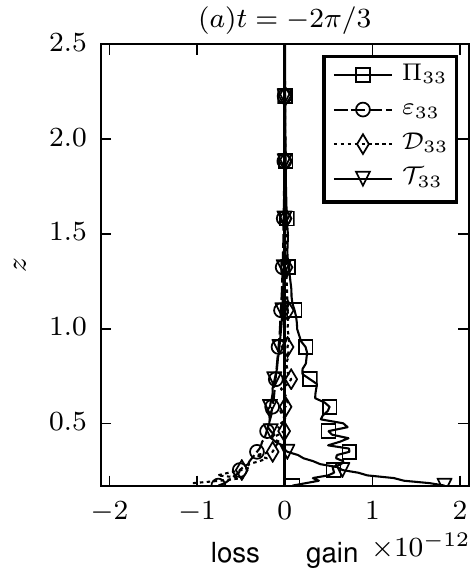}
\includegraphics[]{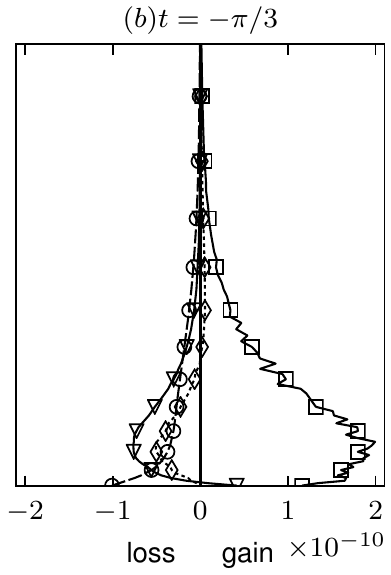}
\includegraphics[]{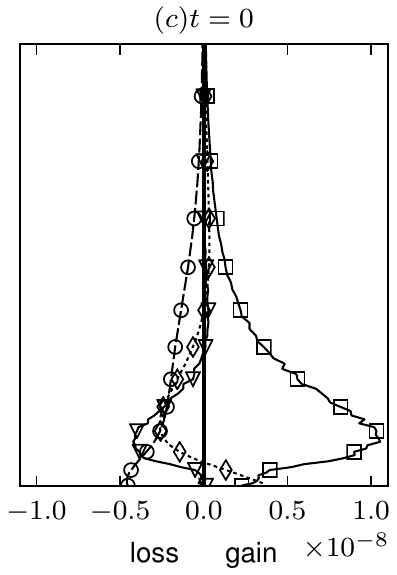}
\caption{ \label{fig:Evv} Energy budget for streamwise-constant vertical fluctuations ($\widetilde {w^\prime}$),  (\ref{eq:Eww}), in case h0.06. ($a$) $t=-2\pi/3$; ($b$) $t=-\pi/3$; ($c$) $t=0$. }
\end{center}
\end{figure}

We have studied thus far streamwise-constant streamwise fluctions, i.e., streaks. Production term $\mathcal P_{11}$ is the manifestation of the lift-up effect driven by streamwise-constant vertical fluctuations $\widetilde {w^\prime}$. Therefore, an essential part of the receptivity process depends on $\widetilde {w^\prime}$. The balance for the kinetic energy of streamwise-constant vertical fluctuations is obtained by multiplying (\ref{eq:wCMom}) with $\widetilde {w^\prime}$ and then ensemble averaging:
\begin{equation}
\begin{split}
\label{eq:Eww}
	\frac{1}{\Rey_\delta}\frac{\partial \langle \widetilde {w^\prime}^2 \rangle}{\partial t}
	&=	\underbrace{- \left \langle \widetilde {w^\prime} \left (\frac{\partial \widetilde{  w^{\prime} v^{\prime}}}{\partial y}+\frac{\partial \widetilde{  w^{\prime} w^{\prime}}}{\partial z}\right)\right \rangle}_{\mathcal T_{33}}
		\underbrace{-\left \langle \widetilde {w^\prime} \frac{\partial \widetilde {p^\prime}}{\partial z}\right \rangle}_{\mathcal \Pi_{33}}\\
	&+\underbrace{\frac{1}{2\Rey_\delta}\frac{\partial^2 \langle \widetilde {w^\prime} \widetilde {w^\prime} \rangle}{\partial z^2}}_{\mathcal D_{33}}
		\underbrace{-\frac{1}{\Rey_\delta}  \left \langle \left(\frac{ \partial \widetilde {w^\prime}}{\partial y}\right)^2+\left(\frac{ \partial \widetilde {w^\prime}}{\partial z}\right)^2\right \rangle}_{\varepsilon_{33}},
	\end{split}
\end{equation}
where $\Pi_{33}$ is the rate of work done by the pressure gradient, and the remaining budget terms with the subscript $33$ are analogous to the terms with subscript $11$ above. There is no production term for the vertical fluctuations. Figure~\ref{fig:Evv} demonstrates the vertical profile of each budget term in case h0.06 at three representative phases $t=-2\pi/3,-\pi/3$ and $0$. At $t=-2\pi/3$, the diffusive transport $\mathcal D_{33}$ and the pressure-gradient work $\Pi_{33}$ are the main contributors to the energy of vertical fluctuations (figure~\ref{fig:Evv}a). At later times, $\Pi_{33}$ is the prevalent gain term. In fully-turbulent shear flows, this term is usually decomposed into redistribution and transport components, among which the redistribution term drives the transfer of energy from streamwise components to cross-stream components \citep{pope2000}. This redistribution mechanism is turned off in streamwise-constant fluctuations, as the streamwise derivative of the streamwise-constant pressure, hence $\Pi_{11}$, vanishes. Therefore, the redistribution and transport can only occur between cross-stream components, and the decomposition does not provide much insight. We need to inspect the instantaneous fields to unravel the origins of $\Pi_{33}$.

\begin{figure}
\begin{center}
\includegraphics[]{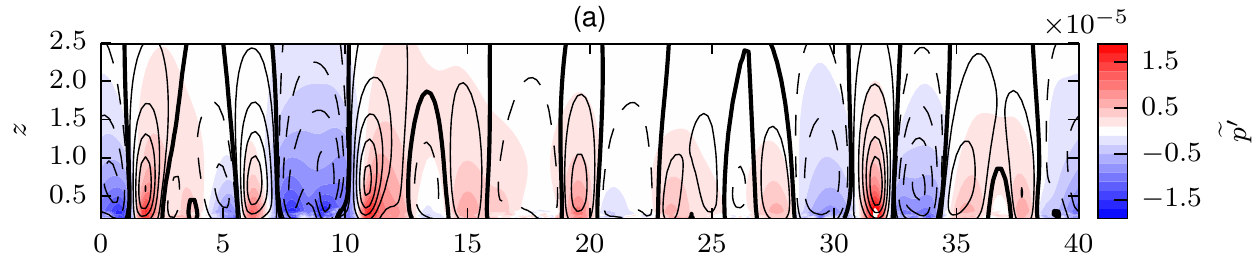}
\includegraphics[]{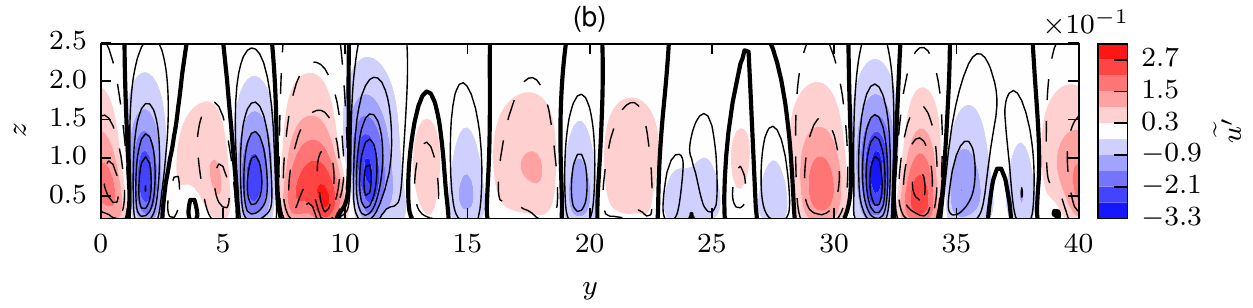}
\caption{ \label{fig:uCpC} Filled contours of $\widetilde {p^\prime}$ ($a$) and $\widetilde {u^\prime}$ ($b$)  in case h0.06 at $t=0$ are plotted with overlaid contour lines of $\widetilde {w^\prime}$. Ten levels of $\widetilde {w^\prime}$ are presented, where negative contours are shown with dashed lines. The thick contour lines show $\widetilde {w^\prime}=0$.}
\end{center}
\end{figure}

Filled-contour distributions of $\widetilde {u^\prime}$ and $\widetilde {p^\prime}$ in case h 0.06 are shown in figure~\ref{fig:uCpC} with overlaid contours of $\widetilde {w^\prime}$ at a representative time instant for the lift-up stage ($t=0$). We observe alternating zones of velocities and pressure separated by zero contours of $\widetilde {w\prime}$ (thick contours).  In these zones, low-speed streaks ($-\widetilde {u^\prime}$) are associated with positive $\widetilde {w^\prime}$ and $\widetilde {p^\prime}$, and high-speed streaks ($\widetilde {u^\prime}$) are associated with negative $\widetilde {w^\prime}$ and $\widetilde {p^\prime}$.
 The filled contours in figure~\ref{fig:uCpC}a further show that the magnitude of pressure in every zone decays upwards from bed. Moreover, the direction of vertical velocity is aligned with negative pressure gradient, i.e., vertical fluctuations are driven down the pressure gradient, hence the positive pressure-gradient work $-\widetilde {w^\prime} \partial \widetilde {p^\prime}/\partial \widetilde z>0$.  This explains the positive correlation between $\widetilde {w^\prime}$ and $\widetilde {p^\prime}$, i.e., in positive pressure zones, upwards decaying pressure drives the vertical velocity upwards, whereas in negative pressure zones, downwards decaying pressure sets a vertical velocity downwards.

  \begin{figure}
\begin{center}
\includegraphics[]{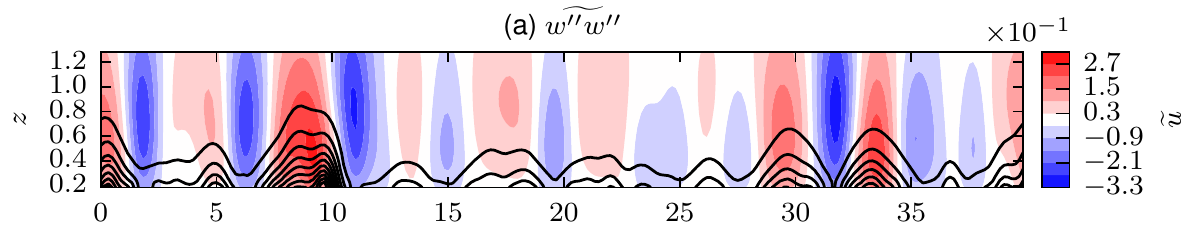}
\includegraphics[]{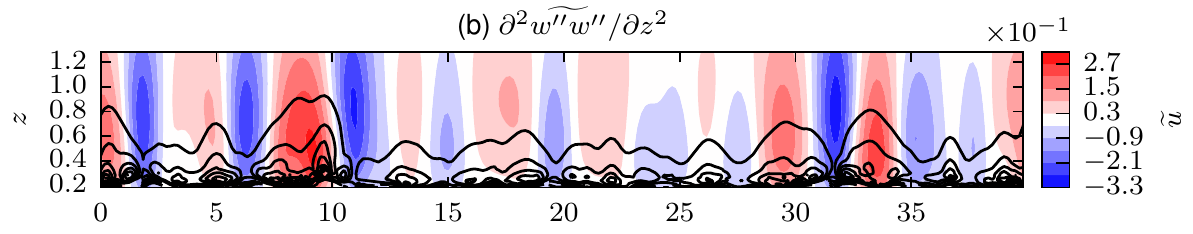}
\includegraphics[]{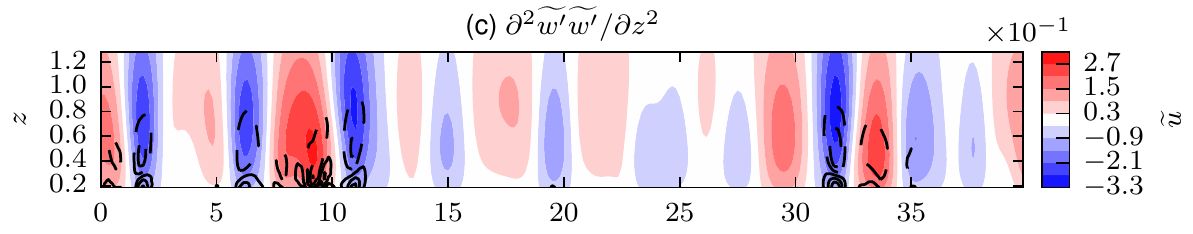}
\includegraphics[]{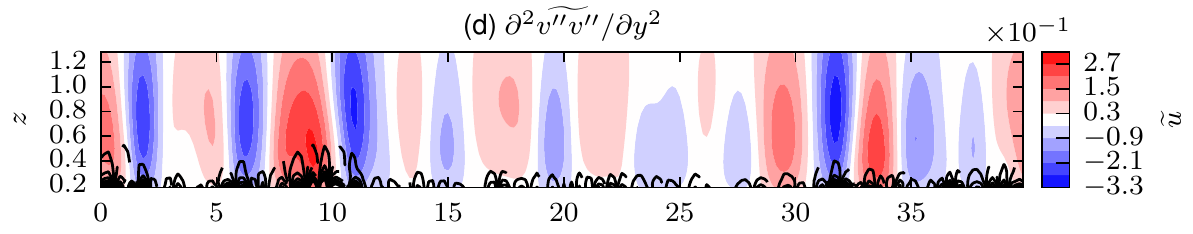}
\includegraphics[]{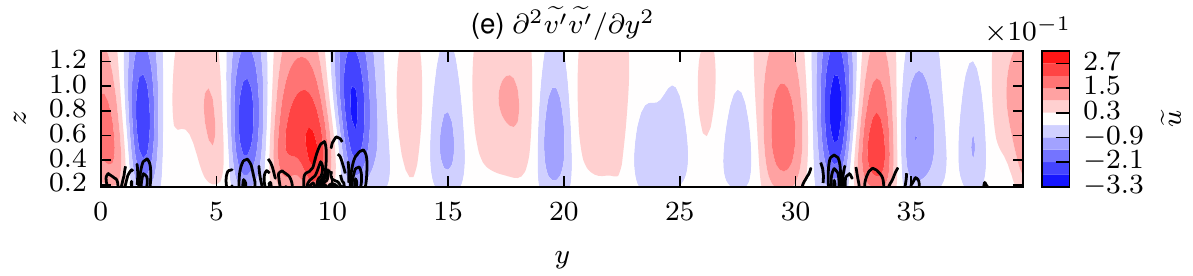}
\caption{ \label{fig:uCvvF} Filled contours of $\widetilde {u^\prime}$ in case h0.06 at $t=0$ are plotted with overlaid line contours of  ($a$) $\widetilde {w^\dprime w^\dprime}$, ($b$) $\partial^2 \widetilde {w^\dprime w^\dprime}/\partial z^2$,  ($c$) $\partial^2 \widetilde {w^\prime} \widetilde {w^\prime}/\partial z^2$, ($d$) $\partial^2 \widetilde {v^\dprime v^\dprime}/\partial y^2$ and ($e$) $\partial^2 \widetilde {v^\prime} \widetilde {v^\prime}/\partial y^2$.  Line contours in ($a$) span 12 levels between $[2.5\times10^{-6},3\times10^{-5}]$, and line contours in ($b$)--($e$) span 12 levels between $[-2.64\times10^{-4},2.64\times10^{-4}]$, where negative contours are shown with dashed lines. }
\end{center}
\end{figure}

  We have seen the alternating $\widetilde {p^\prime}$ zones play a key role in organizing the $\widetilde {w^\prime}$ zones. $\widetilde {p^\prime}$ is forced by second-order variations of second-order terms based on total fluctuation velocities $v^{\prime}$ and $w^{\prime}$, cf.  (\ref{eq:PPE}). Among these forcing terms, $\partial^2 \langle w^{\prime} w^{\prime}\rangle /\partial z^2$ has no effect on spanwise variations in $\widetilde {p^\prime}$, and the cross term between $v^{\prime}$ and $w^{\prime}$ is vanishingly small. The remaining second-order terms  can be decomposed into filtered and residual small-scale components, e.g., $\widetilde {w^{\prime}w^{\prime}}=\widetilde {w^\prime} \widetilde {w^\prime}+\widetilde {w^{\dprime}w^{\dprime}}$. Figures~\ref{fig:uCvvF}($b$--$e$) depict the line contours of the four decomposed forcing terms along with the filled contours of $\widetilde {u^\prime}$  in case h0.06 at instance $t=0$. The term with small-scale vertical fluctuations clearly dominate over other terms  (figures~\ref{fig:uCvvF}b).  The intensive regions of $\partial^2 \widetilde {w^\dprime w^\dprime}/\partial z^2$ are associated with energetic small-scale vertical fluctuations, which are shown in figure~\ref{fig:uCvvF}($a$). Therefore, the residual fluctuations $w^\dprime$ are the essential drivers of $\widetilde {p^\prime}$. These fluctuations are produced when the boundary layer passes over bed topography, and therefore, they scale with the shear imposed at the bed level. In this regard, the energy of small-scale vertical fluctuations are clearly higher in zones of high-speed streaks (positive $\widetilde {u^\prime}$ zones) due to higher shear imposed at the footprints of high-speed streaks. This large-scale amplitude modulation is similar to the one driving inner-outer interactions in turbulent boundary layers \citep{mathis2009large}. Enhanced small-scale energy along high-speed streaks leads to stronger second-order derivatives ($\partial^2 \widetilde {w^\dprime w^\dprime}/\partial z)$ (figure~\ref{fig:uCvvF}$b$), thus stronger forcing of pressure along high-speed streaks. As this forcing is in negative direction (note the negative sign in  (\ref{eq:PPE})), this creates a negative pressure zone along high-speed streaks. Therefore, the  modulation of small-scale fluctuations by large-scale streaks plays a key role in coupling high $\widetilde {u^\prime}$ with low  $\widetilde {p^\prime}$, and vice versa.

      \begin{figure}
\begin{center}
\includegraphics[]{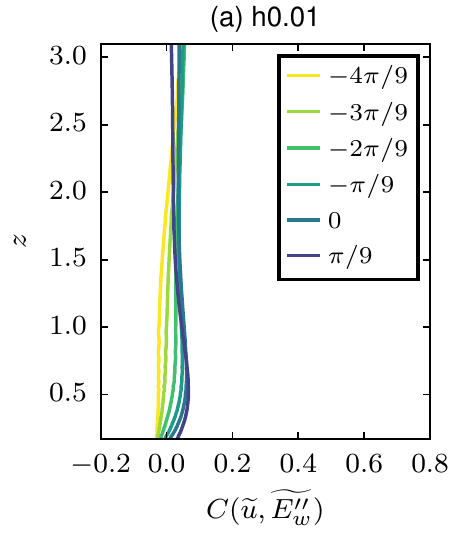}
\includegraphics[]{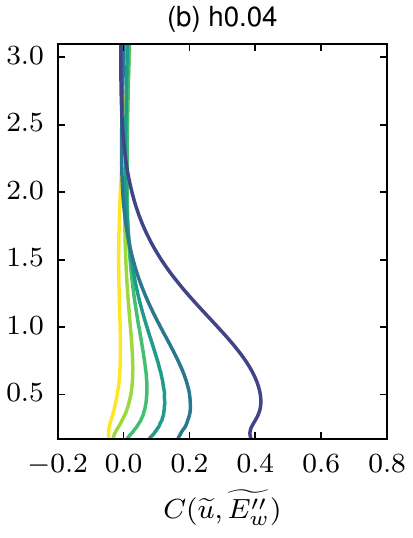}
\includegraphics[]{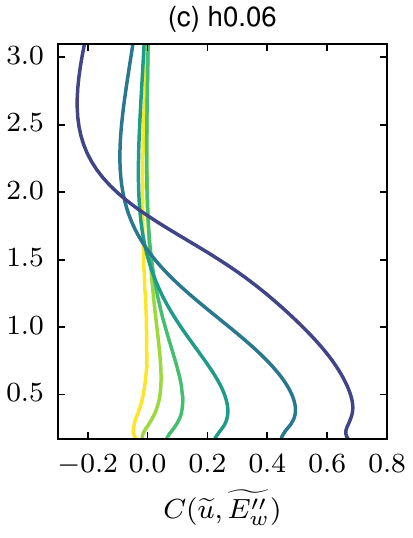}
\caption{ \label{fig:CUEw} Profiles of cross-correlation coefficient $C(\widetilde {u^\prime},  \widetilde {E_w^\dprime})$, where $\widetilde {E_w^\dprime} :=  \widetilde {w^\dprime w^\dprime}$/2. ($a$) h0.01; ($b$) h 0.04; ($c$) h0.06. Color coding is the same for all figures.}
\end{center}
\end{figure}


The effect of large-scale amplitude modulation can be quantified by correlations between large-scale velocity $\widetilde {u^\prime}$ and the energy of small-scale vertical fluctuations  $\widetilde {E_w^\dprime} :=  \widetilde {w^\dprime w^\dprime}/2$, i.e.,
\begin{equation}
\label{eq:C}
C(\widetilde {u^\prime},\widetilde  {E_w^\dprime},z,t)=\frac{\langle \widetilde {u^\prime}(y,z,t) \widetilde {E_w^\dprime}(y,z,t) \rangle}{\sqrt{\langle \widetilde {u^\prime}^2\rangle}\sqrt{ \langle  \widetilde {E_w^\dprime}^2\rangle}}.
\end{equation}
 The vertical profiles of these cross-correlation coefficients are plotted in figure~\ref{fig:CUEw} for cases h0.01, h0.04 and h0.06. $C(\widetilde {u^\prime},  \widetilde {E_w^\dprime})$ increases dramatically with the roughness height. While the amplitude modulation is ineffective in h0.01 ($C(\widetilde {u^\prime},  \widetilde {E_w^\dprime})<0.1$ at all times), it is prevalent in h0.06 with $C(\widetilde {u^\prime},  \widetilde {E_w^\dprime})$ reaching about 70$\%$ correlation  at later times. 

\begin{figure}
\begin{center}
\includegraphics[width=.94\textwidth]{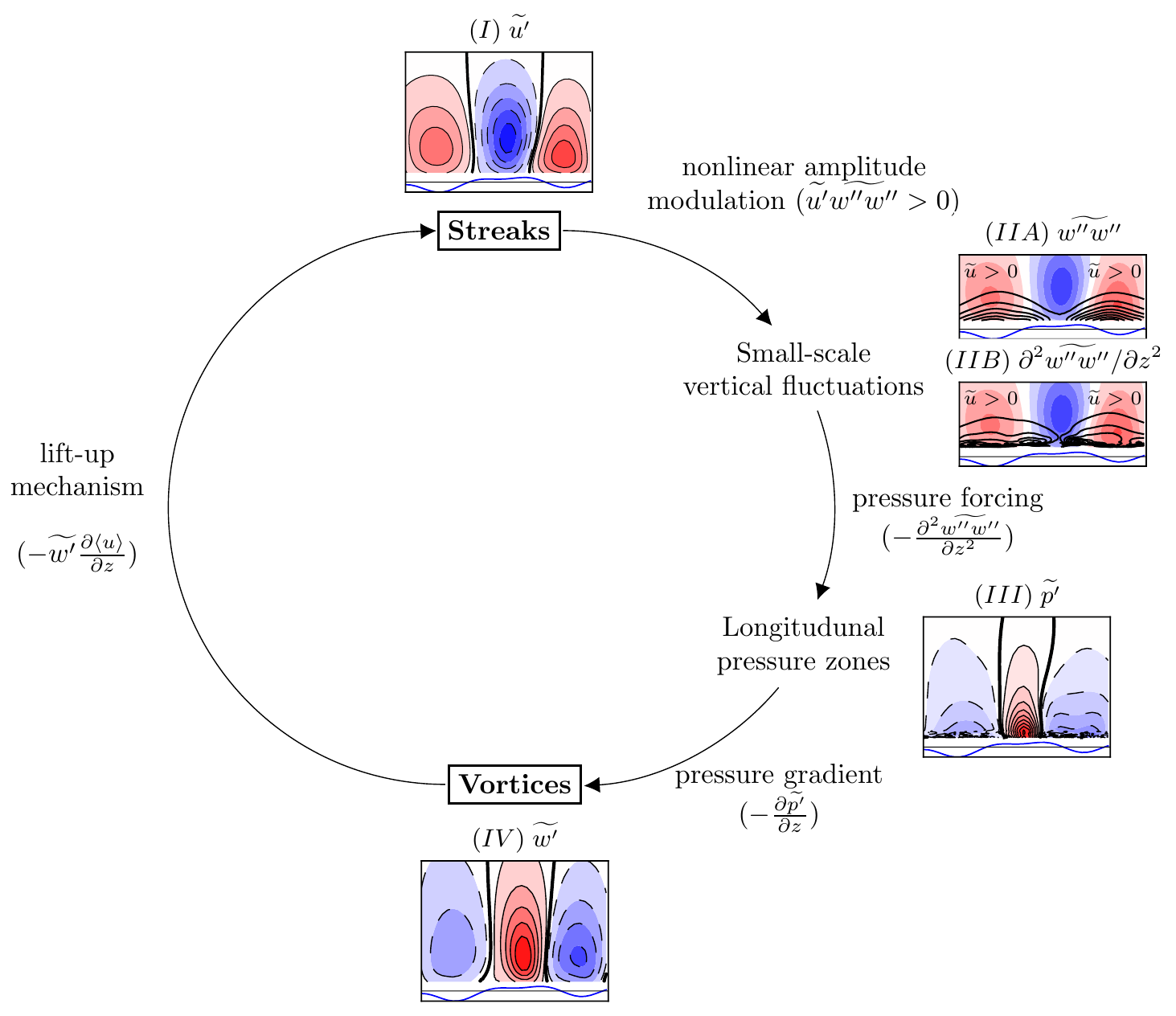}
\caption{ \label{fig:sketch} Positive feedback loop between streamwise-constant streaks and vortices.  The cross-stream components have similar amplitudes, and only the vertical velocity component is considered to represent the vortices. }
\end{center}
\end{figure}

Figure~\ref{fig:sketch} summarizes the second stage of the receptivity process, in which streaks are generated by the lift-up mechanism. This stage is characterized by a positive feedback loop between streamwise-constant streaks  and vortices, i.e., cross-stream components ($\widetilde {v^\dprime}, \widetilde {w^\dprime}$), among which we only consider $\widetilde {w^\dprime}$ for brevity. Streaks ($I$) modulate the small-scale vertical motions, $\widetilde {w^\dprime w^\dprime}$ ($IIA$), whose vertical derivatives ($IIB$) in turn impose alternating zones of high and low streamwise-constant pressure ($III$) aligned with low and high-speed streaks, respectively. The pressure-gradients in these zones induce stronger streamwise-constant vertical velocities, hence vortices ($IV$). Finally, the vortices stir the boundary layer and generate more intense streaks ($IV\rightarrow I$). 

\section{Breakdown stage: transition modes}\label{sec:breakdown}
The receptivity stage was characterised by the dynamics of streamwise-constant perturbations $\widetilde {\vec u^\prime}$. The breakdown stage will be characterized now by the growth of residual perturbations $\vec u^\dprime$ due to primary (orderly transition) or secondary (bypass transition) instabilities.  The paths leading to these instabilities are strongly mediated by streaks. Depending on their amplitude, streaks can damp the growth in regions they occupy, trigger local breakdown by rapidly growing secondary instabilities, or be completely dormant in an orderly transition  scenario \citep{onder_liu_2020}.  We will study these scenarios in this section.

\begin{figure}
\begin{center}
\includegraphics[]{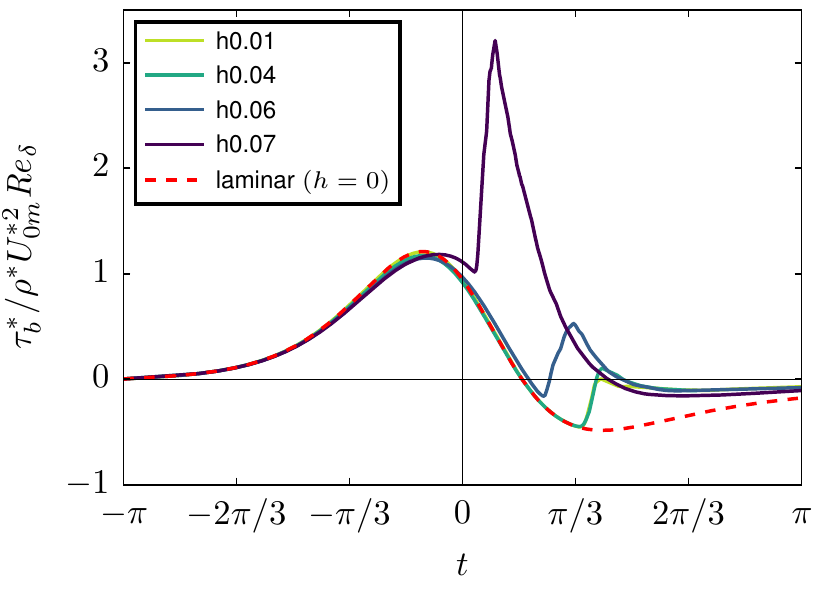}
\caption{  \label{fig:drag} Temporal evolution of mean skin-friction drag for cases h0.01, h0.04, h0.06 and h0.07. Laminar skin-friction drag over flat bed ($h=0$) is also plotted. Skin frictions are normalized by $\rho^* U_{0m}^{*2}\Rey_\delta$. }
\end{center}
\end{figure}

 The time evolution of mean skin-friction drag $\tau_b^*$  is plotted in figure~\ref{fig:drag} for cases h0.01, h0.04, h0.06 and h0.07. In all cases, there is a rapid rise in the skin-friction drag, once the transition sets in. It is clear that the breakdown to turbulence has a much faster timescale compared to wave timescale.  Therefore, during breakdown the streamwise-constant fields have much slower dynamics than the rapidly growing residual perturbations. In this regard, the instantaneous streaky fields represent a new laminar base state  on which the instabilities grow. These base fields are obtained by applying the filter on instantaneous velocity  $\vec u$:
 \begin{equation}
 	\widetilde {\vec u}=(\langle u \rangle,0,0)+(\widetilde {u^\prime},0,0).
 	\label{eq:Us}
 \end{equation}  
We have neglected $\widetilde {v^\prime}$ and $\widetilde {w^\prime}$, as $\|\widetilde {v^\prime}\|\approx \|\widetilde {w^\prime}\| \ll \|\widetilde {u^\prime}\|$ for $\Rey_\delta \gg 1$. 
  
  The growth of secondary perturbations on a streak can be investigated by averaging the small-scale energy over the streamwise direction, i.e., by filtering the small-scale energy: $\widetilde {k^\dprime}(y,z,t)=\widetilde{u_i^{\prime\prime}u_i^{\prime\prime}}/2$.  The instantaneous balance of  $\widetilde {k^\dprime}$ is derived in four steps: (i) filter (\ref{eq:mom}); (ii) subtract the resulting filtered set of equations from (\ref{eq:mom}) to obtain momentum equations for $\vec u^{\dprime}$, (iii)  apply a scalar product between vectorial terms in the resulting momentum equation and $\vec u^{\dprime}$; (iv): filter the resulting energy equations. As a result, we obtain:
\begin{equation}
\frac{1}{\Rey_\delta} \frac{\partial \widetilde {k^\dprime}}{\partial t}+\widetilde \nabla \cdot \widetilde {\vec T^\dprime}= \widetilde{\mathcal P^\dprime} -\widetilde {\varepsilon^\dprime} 
\end{equation}
where $\widetilde{\mathcal P^\dprime}$ represents the small-scale production rate expressed by
\begin{equation}
			\widetilde{\mathcal P^\dprime}= -\widetilde {u^\dprime v^\dprime}\frac{\partial \widetilde u}{\partial y}-\widetilde {u^\dprime w^\dprime}\frac{\partial \widetilde u}{\partial z},
	\label{eq:prod}
\end{equation}
$\widetilde {\varepsilon^\dprime} $ is the dissipative term for streamwise-varying fluctuations
\begin{equation}
	\widetilde {\varepsilon^\dprime}=\frac{2}{\Rey_\delta} \widetilde {s_{ij}^\dprime s_{ij}^\dprime},
\end{equation}
with $s_{ij}^\dprime=1/2 (\partial u_i^\dprime/\partial x_j+\partial u_j^\dprime/\partial x_i)$ being the rate of strain tensor involving small-scale motions, and $\widetilde {\vec T^\dprime}$ contains the turbulent transport terms:
\begin{equation}
	\widetilde {T_i^\dprime}=\frac{1}{2}\widetilde {u_i^\dprime u_j^\dprime u_j^\dprime}+\widetilde{u_i^\dprime p^\dprime}-\frac{2}{\Rey_\delta}\widetilde{u_j^\dprime s_{ij}^\dprime}.
\end{equation}
Among these budget terms we will focus only on the production term in (\ref{eq:prod}), as the instabilities are driven by this term. 

\begin{figure}
\begin{center}
\begin{tabular}{cc}
\begin{tabular}{cc}
     \subfloat[$8\pi/36$]{\includegraphics[]{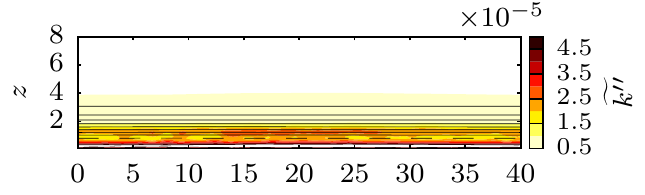}} & 
     \subfloat[$11\pi/36$]{\includegraphics[]{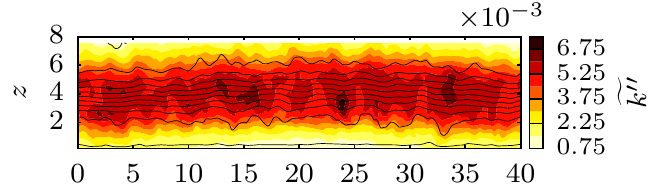}} \\
    \subfloat[$8\pi/36$]{\includegraphics[]{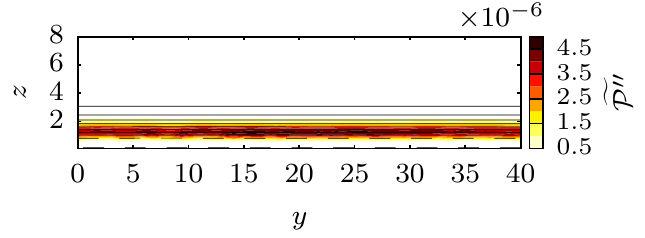}} & 
     \subfloat[$11\pi/36$]{\includegraphics[]{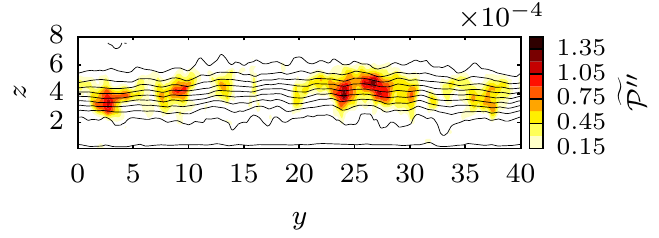}} \\
    \subfloat[$8\pi/36$]{\includegraphics[width=0.49\textwidth]{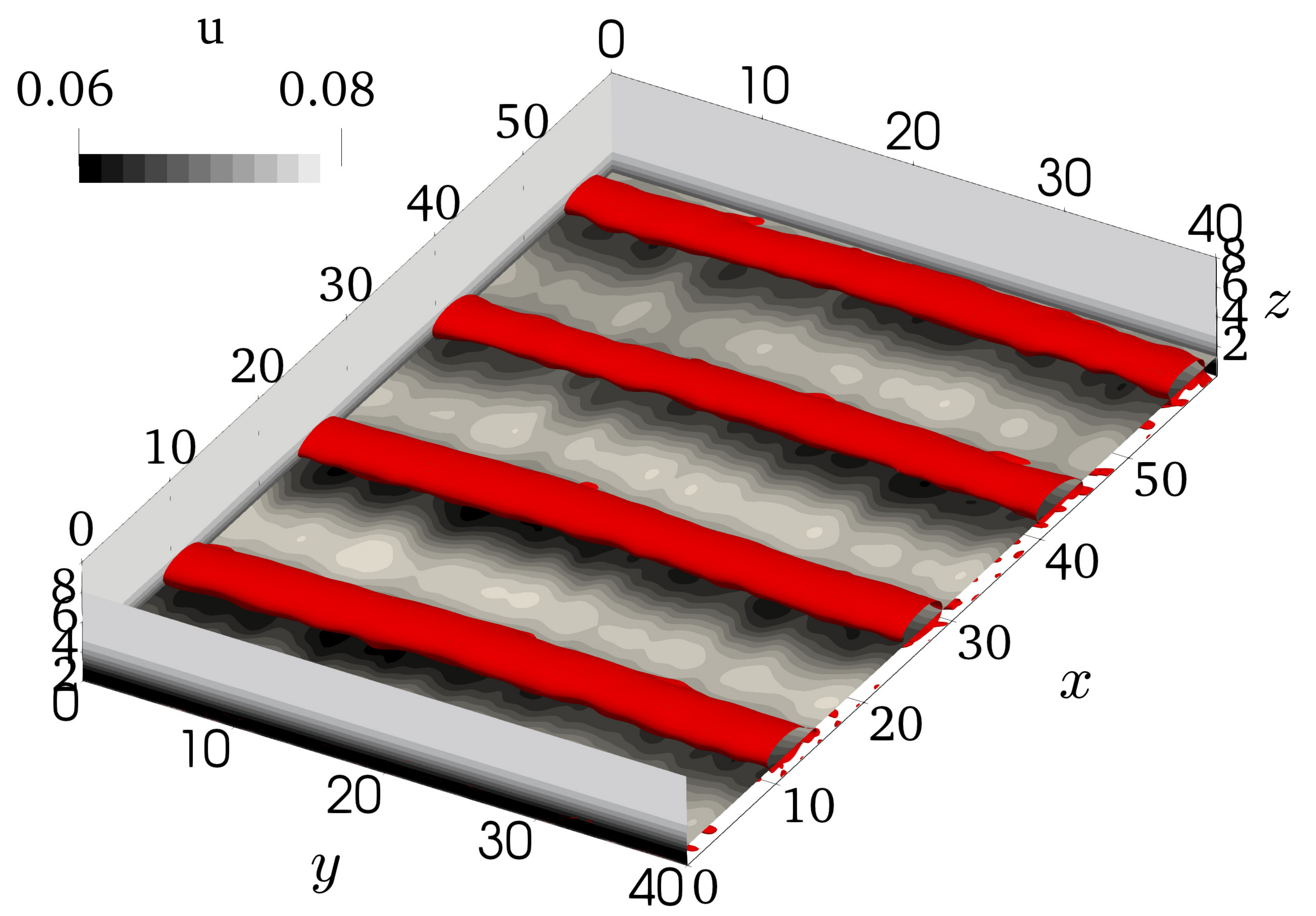}} &
     \subfloat[$11\pi/36$]{\includegraphics[width=0.49\textwidth]{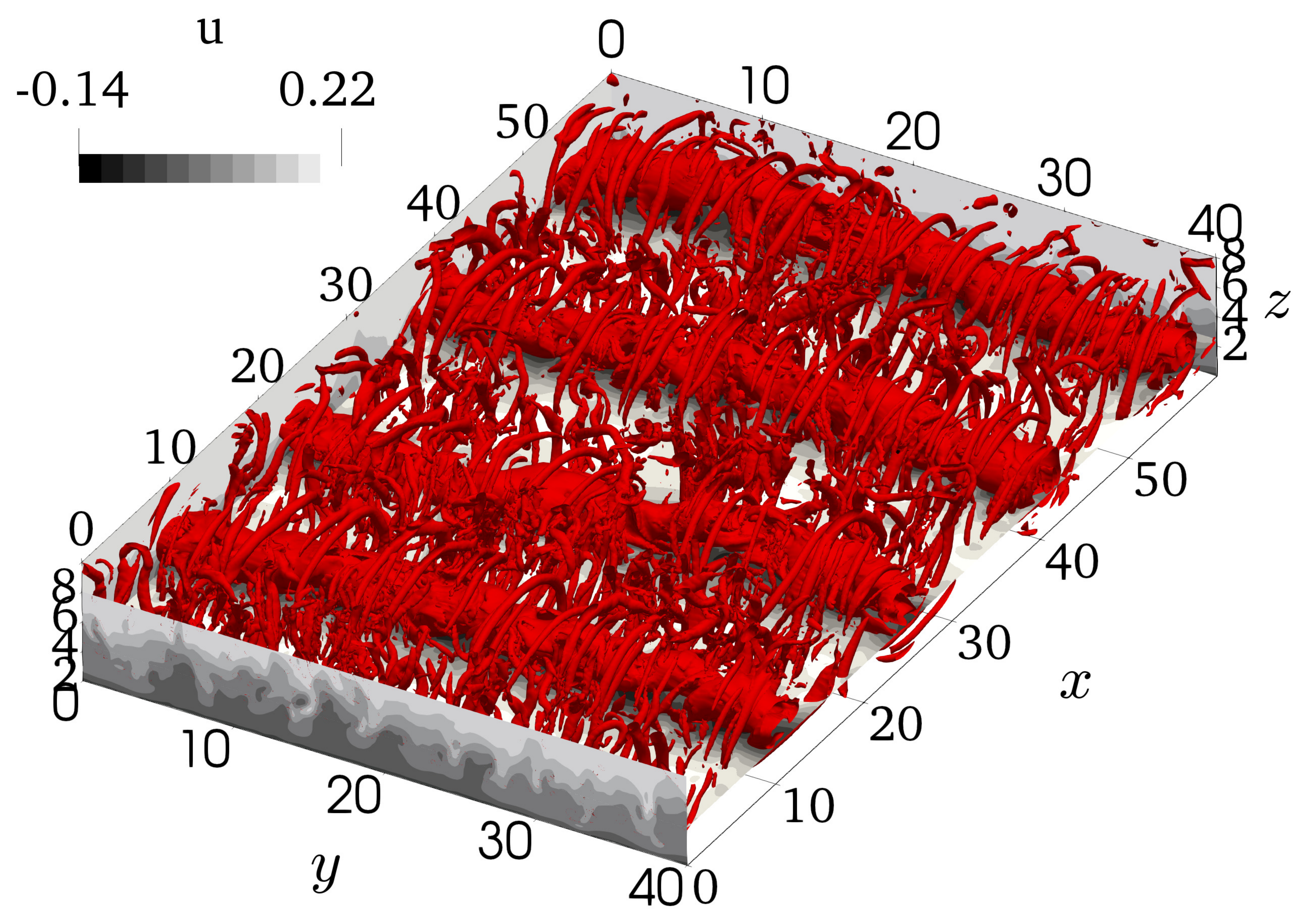}} 
 \end{tabular}
  \end{tabular}
\caption{ Orderly transition  in case~h0.01 via spanwise vortices and their breakdown, ($a,c,e$): $t=8\pi/36$; ($b,d,f$): $t=11\pi/36$. Line contours in ($a$--$d$) show the distribution of $U$ (\ref{eq:Us}) using ten levels. Filled contours show the distribution of the filtered small-scale kinetic energy $\widetilde{k^\dprime}$  ($a,b$) and the production rate $\widetilde{\mathcal P^\dprime}$ ($c,d$).  ($e,f$): Filled contours show the distribution of instantaneous streamwise velocity $u$.   Vortical structures are visualized using a positive isosurface of $Q$. ($e$) $Q=2.5\times 10^{-4}$; ($f$) $Q=8\times 10^{-3}$.\label{fig:Trb0100}}
\end{center}
\end{figure}
 
The transition to turbulence in case h0.01 is shown in figure~\ref{fig:Trb0100} using instantaneous fields, small-scale kinetic energy ($\widetilde {k^\dprime}$) and production rate ($\widetilde{\mathcal P^\dprime}$) at two time instances $t=8\pi/36$ and $11\pi/36$. An orderly transition scenario is observed, in which spanwise vortices emerged (figure~\ref{fig:Trb0100}$e$).  These vortices are visualized using a positive isosurface of Q field, which is the second invariant of the velocity gradient tensor~\citep{Hunt88}:
\begin{equation}
Q=\frac{1}{2}(\Omega_{ij}\Omega_{ij}-s_{ij}s_{ij}),
\label{eq:Q}
\end{equation}
where $\Omega_{ij}=1/2 (\partial  u_i /\partial x_j- \partial  u_j/\partial x_i  )$ is the rate of spin tensor, and $s_{ij}=1/2 (\partial  u_i /\partial x_j+\partial  u_j/\partial x_i)$ is the rate of strain tensor. The vortices have a constant streamwise spacing of $\lambda_x=L_x/4=15$. This spacing perfectly matches the wavelength of the most linearly unstable modes at $\Rey_\delta=2000$ (cf. figure~12a in \cite{onder_liu_2020}). The production is localized at $z\approx 1.2$ where the instability is generated (figure~\ref{fig:Trb0100}$c$). At $11\pi/36$, the spanwise vortices has become two orders of magnitude more energetic (figure~\ref{fig:Trb0100}$b$), and are elevated further into free-stream. At this phase, the coherent vortices are turbulent structures that are in the process of breakdown to chaotic small-scale motions, cf. figure~\ref{fig:Trb0100}($f$). In this global transition scenario, streaks play no role, as they are very weak. 

\begin{figure}
\begin{center}
\begin{tabular}{cc}
     \subfloat[$19/90\pi$]{\includegraphics[]{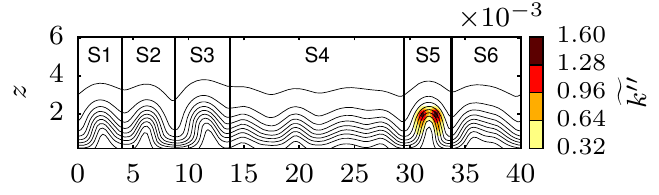}} & 
     \subfloat[$22/90\pi$]{\includegraphics[]{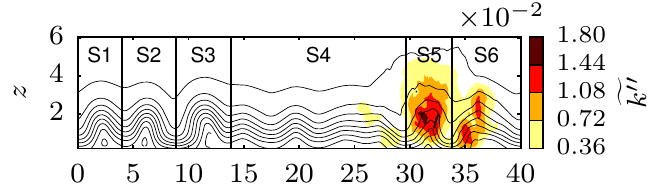}} \\
    \subfloat[$19/90\pi$]{\includegraphics[]{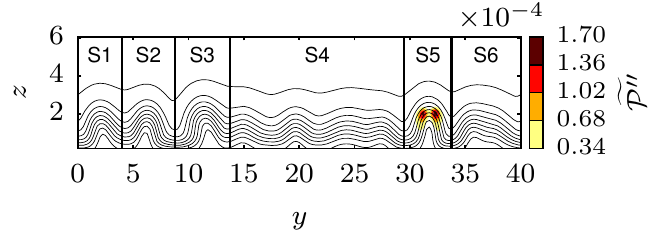}} & 
     \subfloat[$22/90\pi$]{\includegraphics[]{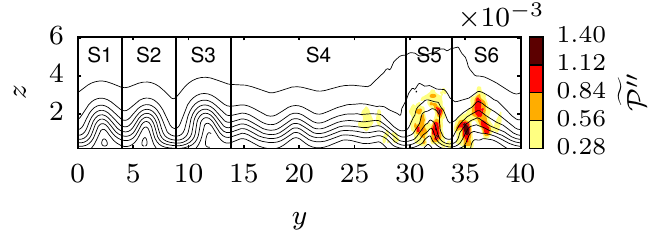}} \\
    \subfloat[$19/90\pi$]{\includegraphics[width=0.49\textwidth]{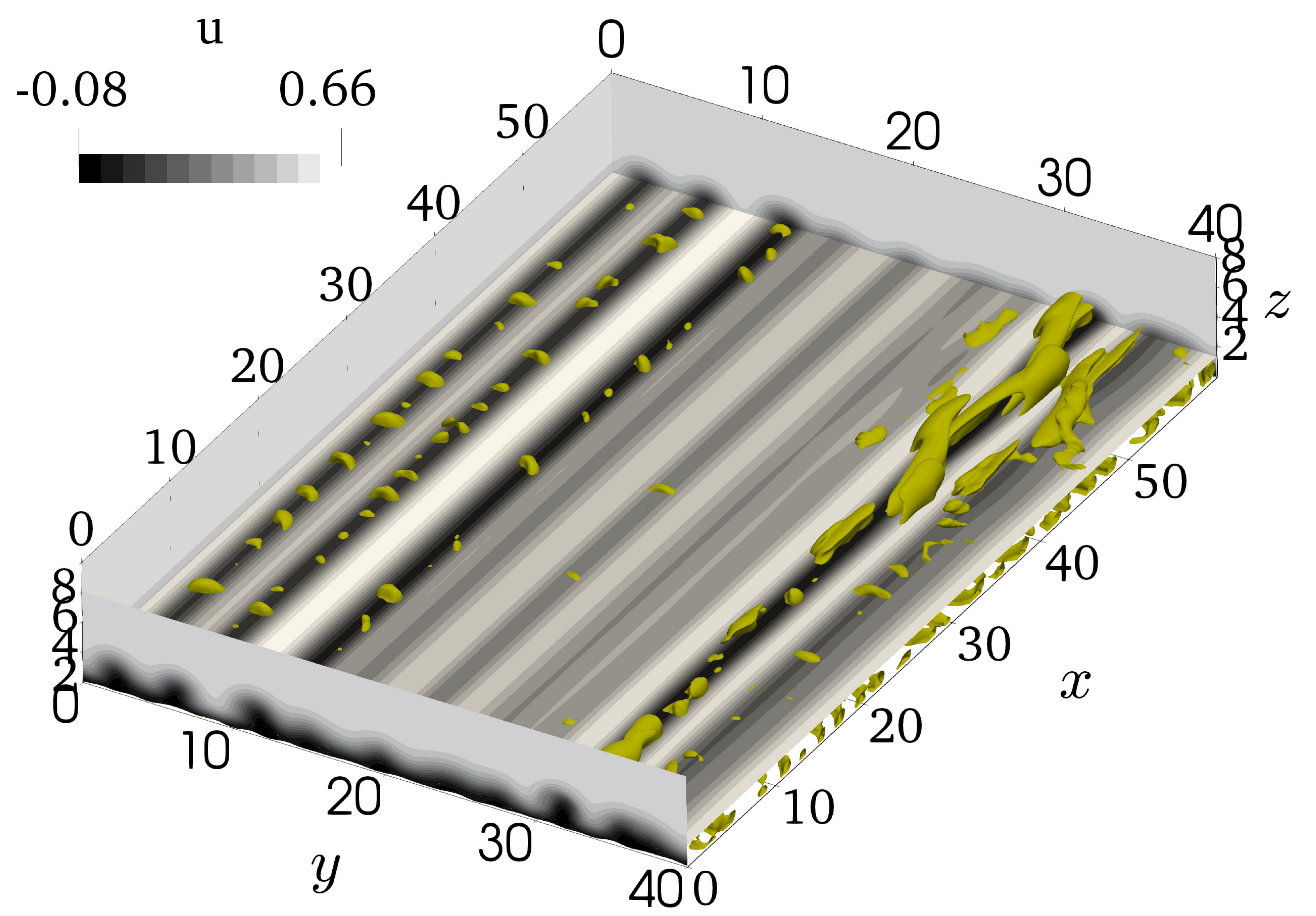}} &
     \subfloat[$22/90\pi$]{\includegraphics[width=0.49\textwidth]{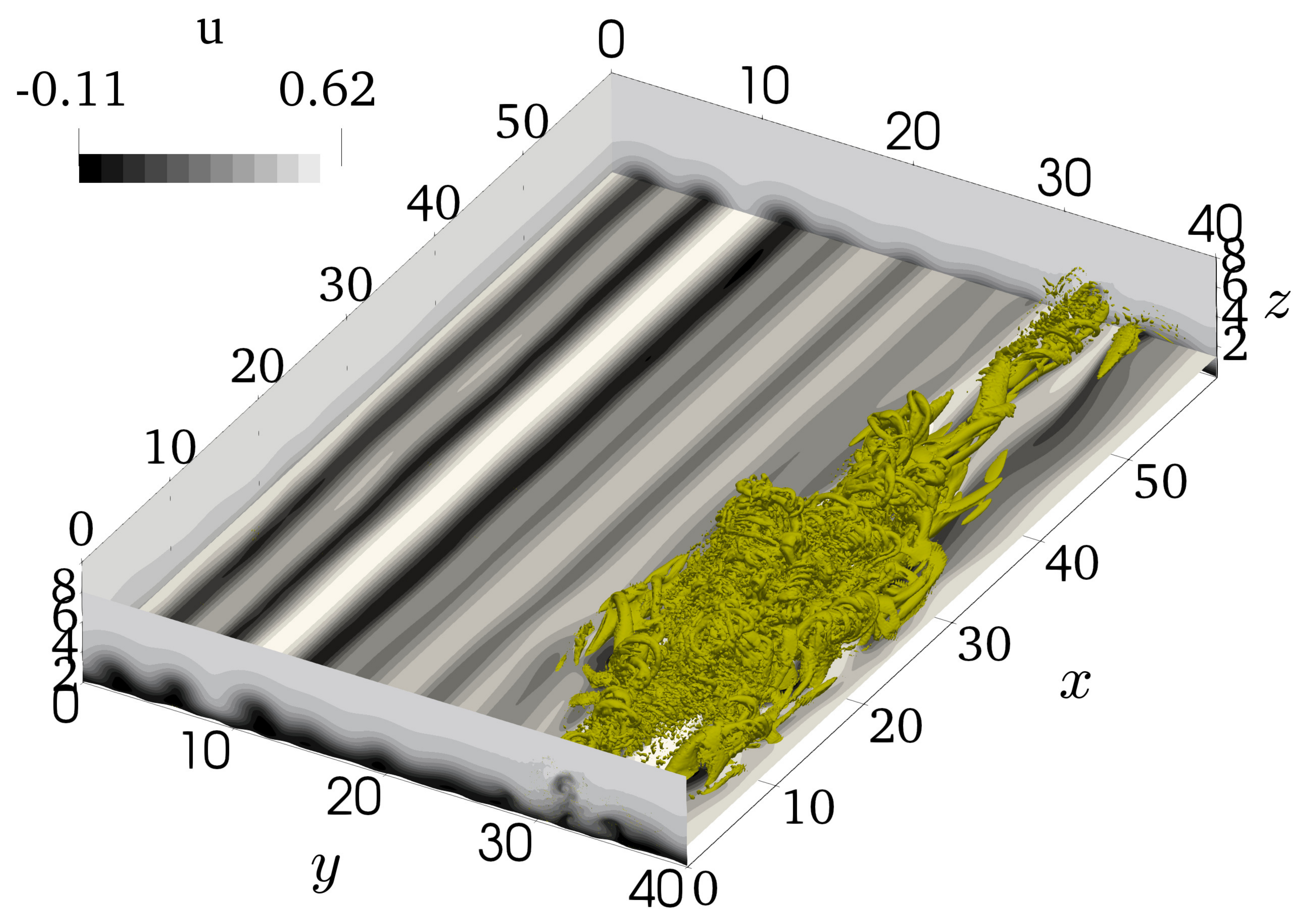}} 
 \end{tabular}
\caption{  Bypass transition in case~h0.06 via a local streak instability. ($a,c,e$): $t=19\pi/90$; ($b,d,f$): $t=22\pi/90$. Line contours in (a-d) show the distribution of $U$ using ten levels.  Filled contours show the distribution of the filtered small-scale kinetic energy $\widetilde{k^\dprime}$  ($a,b$) and the production rate $\widetilde{\mathcal P^\dprime}$ ($c,d$). ($e,f$): Filled contours show the distribution of instantaneous streamwise velocity $u$.   Vortical structures are visualized using a positive isosurface of $Q$. ($e$) $Q=2.5\times 10^{-4}$; ($f$) $Q=8\times 10^{-3}$.  \label{fig:Trb0600}}
\end{center}
\end{figure}

The transition to turbulence in case h0.06 is demonstrated in figure~\ref{fig:Trb0600} using instantaneous data at times $t=19\pi/90$ and $22\pi/90$. The boundary layer is corrugated along its span by  highly elevated streaks (figure~\ref{fig:Trb0600}$a$). We observe five such streaks (S1--S3, S5 and S6 in figure ~\ref{fig:Trb0600}$a$--$d$) and a relatively flat region at the center with weak streaks (S4). Unlike the transition in h0.01, the transition in h0.06 is of local nature and is initiated by a sinuous instability taking place on streak S5, cf. figure~\ref{fig:Trb0600}($e$). Both production and kinetic energy concentrate at an outer layer at $z\approx 2$ marking the location of the critical layer of the instability (figure~\ref{fig:Trb0600}$a,c$). The sinuous nature and strongly elevated critical layer suggest that this instability is an instance of outer-streak instabilities cited by \cite{onder_liu_2020}. The outer instabilities have very high growth rates and rapidly lead to bypass transition. This is observed in figure~\ref{fig:Trb0600}f, where streak S5 broke down into a turbulent spot. Turbulence is contained in this region and the rest of the boundary layer is still laminar. In later times (not shown here), the turbulent spot spreads to the whole computational domain and the breakdown to turbulence is completed. The instability waves did not emerge in this case.

\begin{figure}
\begin{flushleft}
\begin{tabular}{ll}
     \subfloat[$27\pi/90$]{\includegraphics[]{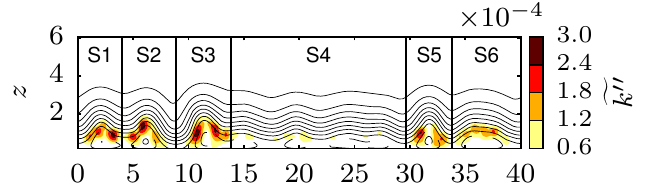}} & 
     \subfloat[$30\pi/90$]{\includegraphics[]{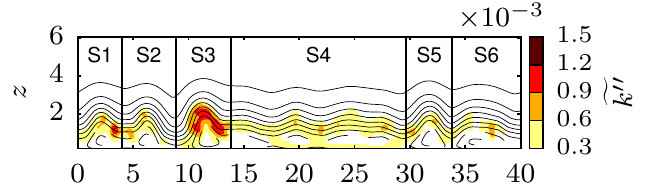}} \\
          \subfloat[$27\pi/90$]{\includegraphics[]{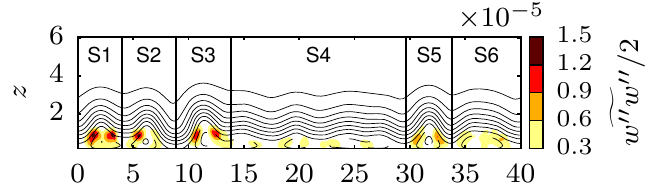}} & 
     \subfloat[$30\pi/90$]{\includegraphics[]{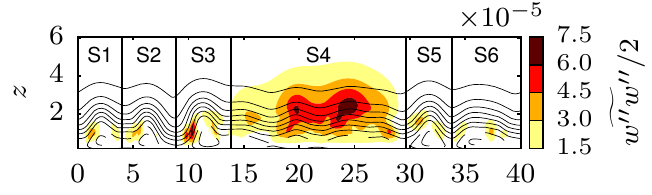}} \\
    \subfloat[$27\pi/90$]{\includegraphics[]{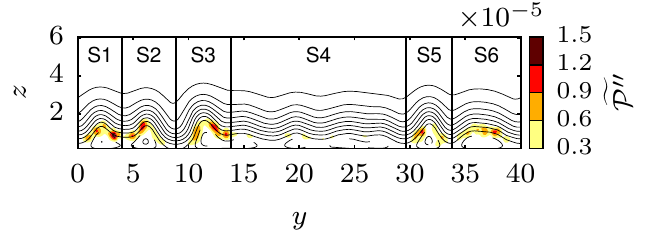}} & 
     \subfloat[$30\pi/90$]{\includegraphics[]{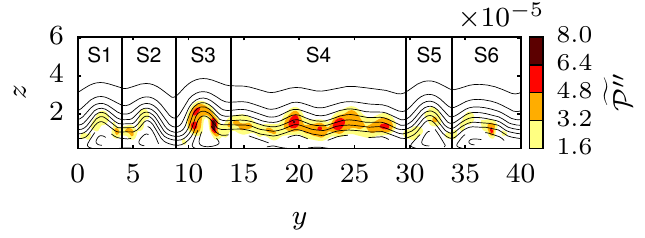}} \\
    \subfloat[$27\pi/90$]{\includegraphics[width=0.49\textwidth]{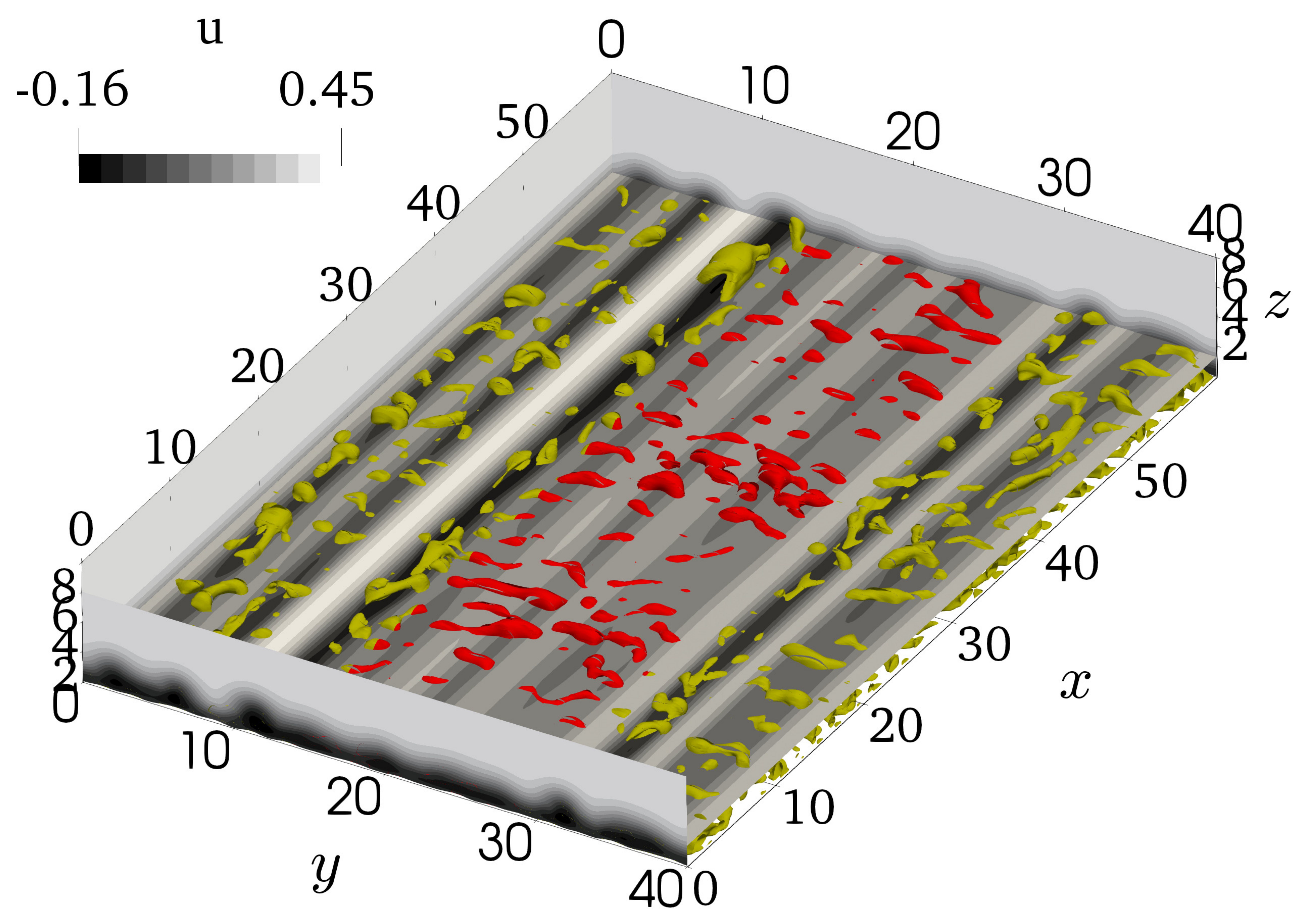}} &
     \subfloat[$30\pi/90$]{\includegraphics[width=0.49\textwidth]{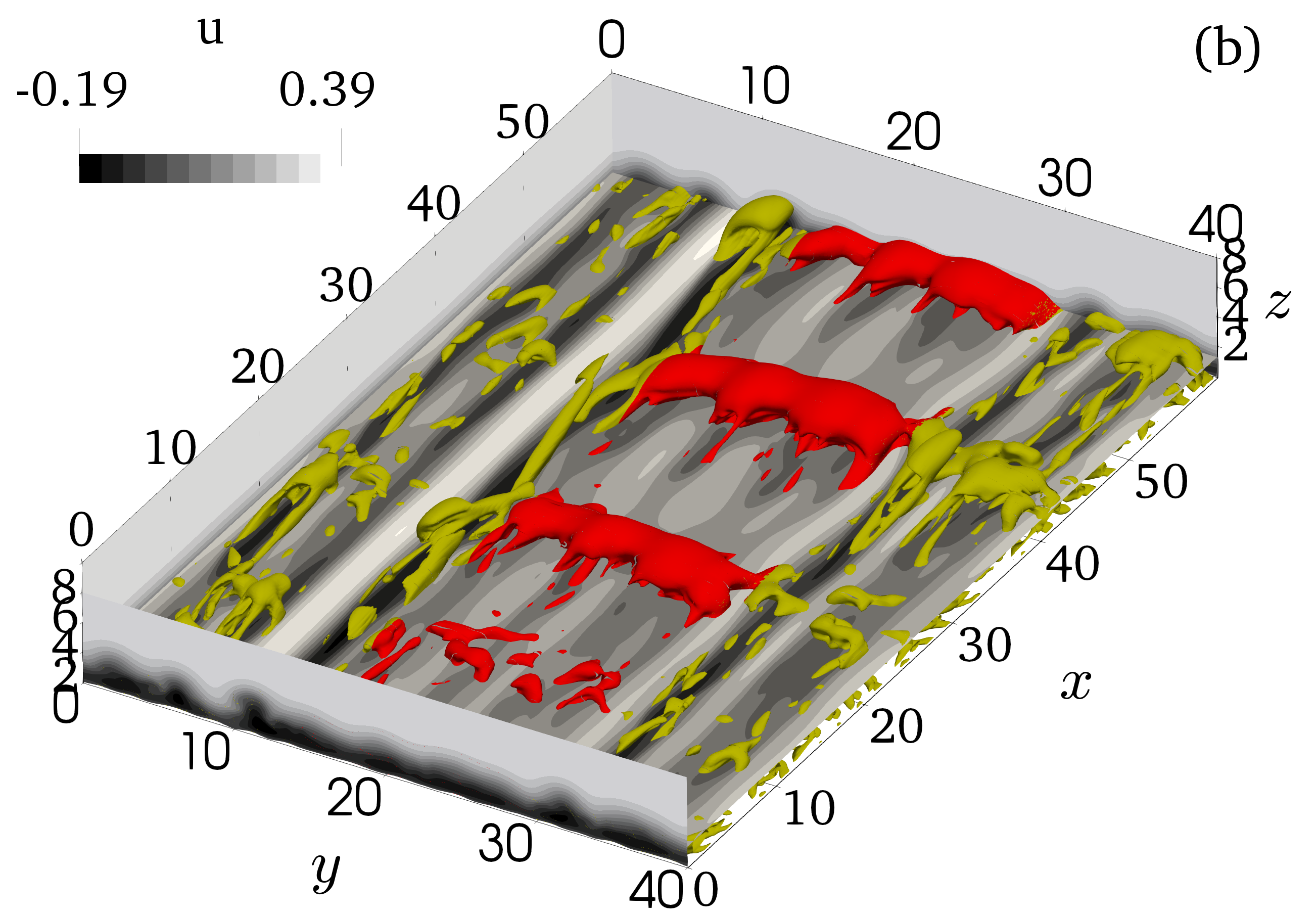}} 
 \end{tabular}
\end{flushleft}
\caption{Mixed transition in case~h0.055. ($a,c,e$): $t=27\pi/90$; ($b,d,f$): $t=30\pi/90$. Line contours in ($a$--$f$) show the distribution of $U$ using ten levels.  Filled contours show the distribution of the filtered small-scale kinetic energy $\widetilde{k^\dprime}$  ($a,b$), the  vertical energy $\widetilde{w^\dprime w^\dprime}/2$  ($c,d$), and the production rate $\widetilde{\mathcal P^\dprime}$ ($e,f$).  ($g,h$): Filled contours show the distribution of instantaneous streamwise velocity $u$.   Vortical structures are visualized using a positive isosurface of $Q$. Streaky regions S1-S3, S5, S6 are colored in yellow, and S4 is colored in red. ($g$) $Q=2.5\times 10^{-4}$; ($h$) $Q=5\times 10^{-4}$.  \label{fig:Trb0550}}
\end{figure}

\begin{figure}
\begin{center}
\includegraphics{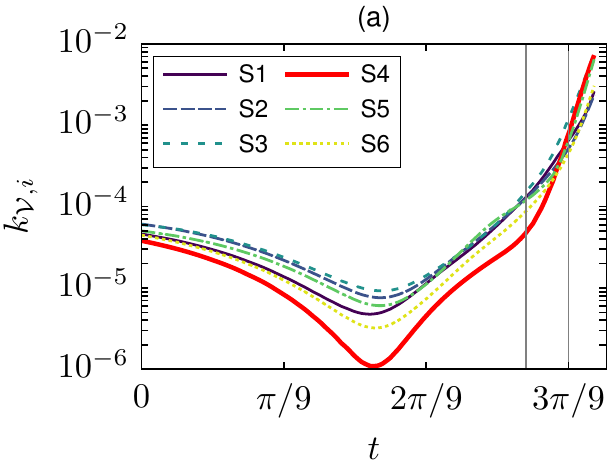}
\includegraphics{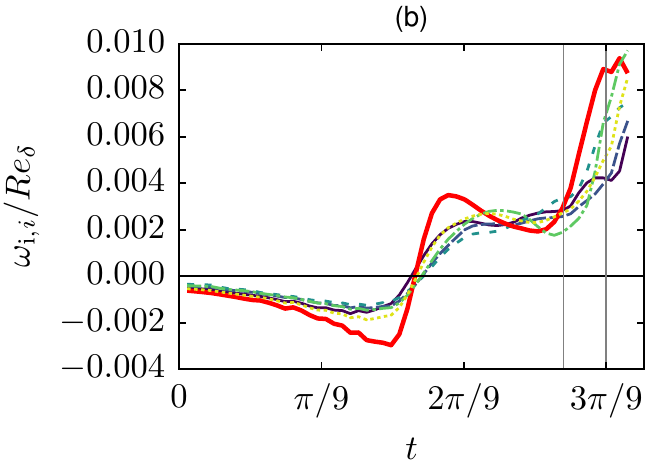}
\caption{Temporal variations of: ($a$) integrated small-scale kinetic energies (\ref{eq:kv}), ($b$) the growth rates  (\ref{eq:omegai}), in regions S1-S6 in case~h0.055. The vertical lines show the instances $t=27\pi/90$ and $30\pi/90$, for which the instantenous fields are shown in figure~\ref{fig:Trb0550}. \label{fig:kL}}
\end{center}
\end{figure}

A mixed transition is demonstrated in figure~\ref{fig:Trb0550}, which occurs in case h0.055. Figure~\ref{fig:kL}a further shows the length-normalized kinetic energy in each subregion (S1--S6),
\begin{equation}
\label{eq:kv}
k_{\mathcal V,i}(t)=\frac{1}{2L_{y,i}}\iint_{A_i} \widetilde{u_i^\dprime u_i^\dprime} (y,z,t) \mathrm d y \mathrm d z,
\end{equation}
where $A_i$ and $L_{y,i}$ are the area and the spanwise length of the subregion, respectively. Figure~\ref{fig:kL}b additionally plots the growth rates of energy in each subregion:
\begin{equation}
\label{eq:omegai}
\omega_{\ii,i}(t)=\frac{1}{2}\frac{\mathrm d\log{k_{\mathcal V,i}}}{\mathrm d t}.
\end{equation}
Small-scale kinetic energy, the vertical kinetic energy and the production rate at $t=27\pi/90$ are plotted in figures~\ref{fig:Trb0550}($a$),~\ref{fig:Trb0550}($c$) and \ref{fig:Trb0550}($e$), respectively.  In h0.055, streaks are slightly weaker compared to those in h0.06, and none of them develops rapidly growing outer instabilities.  Streaks S1-S3, S5 and S6 all developed instabilities on inner shear layers close to the bed. These inner instabilities have a slow growth rate of $\omega_\ii/\Rey_\delta \approx 3\times10^{-3}$, cf. figure~\ref{fig:kL}($b$). These slow growth rates are consistent with the inner-streak instabilities analyzed by \citeauthor {onder_liu_2020} (cf. figure~12$b$ in \cite{onder_liu_2020}). Due to their slow growth, the observed inner instabilities did not lead to any intense turbulent structure yet (figure~\ref{fig:Trb0550}$g$). In contrast to streaky regions, the region S4 is relatively quiet at $t=27\pi/90$, and fluctuations are very weak (cf. S4 in figures~\ref{fig:Trb0550}$a$ and \ref{fig:kL}$a$). This condition changes abruptly due to spontaneously emerging instability waves, and we observe later at $t=30\pi/90$ coherent spanwise vortices with streamwise spacing of $\lambda_x=15$ in S4, cf. red Q-isosurfaces in figure~\ref{fig:Trb0550}($h$). The quasi two-dimensional instability taking place in S4 has a higher growth rate than inner-streak instabilities with values in the range $\omega_{\ii,4}/\Rey_\delta \approx 10^{-2}$ (figure~\ref{fig:kL}$b$). Compared to the local transitional features in inner-streak instabilities, spanwise coherent vortices are more global structures, which occupy the whole boundary-layer thickness. This can be seen in the distribution of vertical kinetic energy in figure~\ref{fig:Trb0550}($d$). Filled contours have spread everywhere in the boundary layer in S4, where the vertical perturbations in other streak regions are still localized around the critical layer of the instability. At later times, the quasi-two-dimensional instability in S4 spreads to neighboring regions S3 and S5, and due to rapid growth of this instability the energies in these regions are significantly higher, e.g. $k_{\mathcal V,i}$ of S3-S5 in the last data point in figure~\ref{fig:kL}($a$) is about four times of that of S1, S2, and S6. 

\section{Conclusions}\label{sec:conclusion}
The present work studies the transition to turbulence in a bottom boundary layer developing over random bottom topography beneath a solitary wave.  The bed is composed of the superposition of wave-like undulations with random amplitude. A set of direct numerical simulations with various roughness levels is conducted, in which the flow around bed corrugations is well resolved by using a coordinate transformation. A relatively high Reynolds number ($\Rey_\delta=2000$) is selected to allow a wide range of transition scenarios.  

In the first part, we analysed the receptivity of the boundary layer flow to perturbations introduced by bottom topography. The boundary layer responds to these broadband perturbations selectively and develops streamwise-elongated streaks, which occupy the whole streamwise extent of the periodic computational domain. These streamwise-constant streaks are generated by two successive physical mechanisms. Initially, when the free-stream velocity is weak, the interaction of the boundary layer with the topography of the bed is linear. To this end, streamwise-constant modes of the topography induce streamwise-constant streaks by diffusive transport, where high- and low-speed streaks are positioned on the depressed and elevated regions, respectively. At later phases, the boundary-layer shear becomes stronger and its interaction with streamwise-constant vertical velocity further amplifies the streaks. Physically, the amplification is driven by the convective transport known as the lift-up mechanism, in which streamwise-constant vortices stir the mean flow. This second stage of streak amplification is characterised by nonlinear feedback loops between streamwise-constant streaks and vortices. When the streaks reach high amplitudes, they begin to modulate the small-scale perturbations at their footprints. These small-scale fluctuations, the vertical component precisely, in turn force streamwise-constant pressure field and create low- and high-pressure zones along high- and low-speed streaks, respectively. The gradients in these pressure zones drive the streamwise-constant vertical velocity, and generate stronger vortices. Stronger vortices in turn generates even stronger streaks via the lift-up mechanism, which completes the positive feedback loop. The consequence of this nonlinear feedback loop is evident in the scaling of streak amplitudes with the roughness height. For instance, at the start of the APG stage, the case with $h=0.06$ has about 85 times stronger streaks than the case with $h=0.01$.

The transition path in the breakdown stage heavily depends on the amplitude of streaks in the respective subregions of the boundary layer. In this regard, three different scenarios are observed: (i) when the streaks are weak the flow goes through orderly transition initiated by  two-dimensional instability waves; (ii) inner-streak instabilities are observed in the regions with moderate amplitude streaks; (iii) outer-streak instabilities are observed in the regions with high-amplitude streaks. Consistent with the previous analysis by authors \citep{onder_liu_2020}, inner-streak instabilities have slower growth rates than primary modal instabilities, and the transition to turbulence is delayed in the regions occupied by moderate-amplitude streaks. Therefore, the current work confirms the stabilizing role of moderate-amplitude streaks. In contrast to inner instabilities, outer  instabilities grow very fast on highly elevated streaks. Turbulent spots are nucleated in these regions, and a bypass-transition scenario is initiated. 

An essential element of the transition over randomly rough topography is the interaction between different transition modes growing in exclusive parts of the domain. We have presented an instance of this phenomenon, in which two-dimensional instabilities and inner-streak instabilities grew separately and eventually interact.  However, many additional scenarios are possible that would require larger domains to study. Larger domains will allow much longer wavelengths in the topography, hence larger scale modulations in the bed elevation. Such modulations can lead to boundary layers hosting a wide spectrum of streak amplitudes. Consequently, bypass, orderly or damped transition scenarios can be initiated separately, and spanwise vortices and turbulent spots can coexist and interact as in the experiments of \cite{Sumer:2010ce}. Such mixed scenarios can occur frequently in Nature over inhomogeneous seabeds. Further study on their dynamics is required to gain a more global perspective on roughness-induced transition in wave boundary layers. 

\section*{Acknowledgements}
The research reported here has been supported by a Tier~2 grant from Ministry of Education of Singapore to National University of Singapore. The computational work for this article was fully performed on resources of the National Supercomputing Centre, Singapore (https://www.nscc.sg). 
\section*{Declaration of interests}
The authors report no conflict of interest.

\bibliographystyle{jfm}
\bibliography{allpapers}

\end{document}